\documentclass[twocolumn]{aastex631}

\shorttitle{Thacher Observatory}
\shortauthors{Swift et al.}
\graphicspath{{./}{figures/}}

\begin{document}

\title{The Renovated Thacher Observatory and First Science Results}

\author[0000-0002-9486-818X]{Jonathan J. Swift}
\affiliation{The Thacher School, 5025 Thacher Rd., Ojai, CA 93023, USA}
\correspondingauthor{J. J. Swift: jswift@thacher.org}

\author[0000-0001-9334-5531]{Karina Andersen}
\affiliation{University of Colorado Boulder, 2055 Regent Dr., Boulder, CO 80305, USA}

\author[0000-0002-5557-7574]{Toby Arculli}
\affiliation{Pomona College, 170 E 6th Street, Claremont, CA 91711 USA}

\author{Oakley Browning}
\affiliation{The Thacher School, 5025 Thacher Rd., Ojai, CA 93023, USA}

\author{Jeffrey Ding}
\affiliation{The Thacher School, 5025 Thacher Rd., Ojai, CA 93023, USA}

\author{Nick Edwards}
\affiliation{The Thacher School, 5025 Thacher Rd., Ojai, CA 93023, USA}

\author{Tom\'{a}s Fanning}
\affiliation{The Thacher School, 5025 Thacher Rd., Ojai, CA 93023, USA}

\author{John Geyer}
\affiliation{The Thacher School, 5025 Thacher Rd., Ojai, CA 93023, USA}

\author{Grace Huber}
\affiliation{The Thacher School, 5025 Thacher Rd., Ojai, CA 93023, USA}

\author{Dylan Jin-Ngo}
\affiliation{The Thacher School, 5025 Thacher Rd., Ojai, CA 93023, USA}

\author{Ben Kelliher}
\affiliation{The Thacher School, 5025 Thacher Rd., Ojai, CA 93023, USA}

\author{Colin Kirkpatrick}
\affiliation{The Thacher School, 5025 Thacher Rd., Ojai, CA 93023, USA}

\author[0000-0002-6476-5331]{Liam Kirkpatrick}
\affiliation{Dartmouth College, 78 College Street, Hanover, NH 03755, USA}

\author{Douglas Klink III}
\affiliation{The Thacher School, 5025 Thacher Rd., Ojai, CA 93023, USA}

\author{Connor Lavine}
\affiliation{The Thacher School, 5025 Thacher Rd., Ojai, CA 93023, USA}

\author{George Lawrence}
\affiliation{The Thacher School, 5025 Thacher Rd., Ojai, CA 93023, USA}

\author{Yousef Lawrence}
\affiliation{University of Chicago, 5801 South Ellis Ave, Chicago, IL 60637, USA}

\author[0000-0002-6604-8838]{Feng Lin Cyrus Leung}
\affiliation{The Thacher School, 5025 Thacher Rd., Ojai, CA 93023, USA}

\author{Julien Luebbers}
\affiliation{The Thacher School, 5025 Thacher Rd., Ojai, CA 93023, USA}

\author[0000-0001-6145-5859]{Justin Myles}
\affiliation{Department of Physics, Stanford University, 382 Via Pueblo Mall, Stanford, CA 94305, USA}

\author[0000-0003-4852-6485]{Theo J. O'Neill}
\affiliation{Department of Astronomy, University of Virginia, Charlottesville, Virginia 22904, USA}
\affiliation{The Thacher School, 5025 Thacher Rd., Ojai, CA 93023, USA}

\author[0000-0001-6052-1973]{Jaime Osuna}
\affiliation{The Thacher School, 5025 Thacher Rd., Ojai, CA 93023, USA}

\author{Peter Phipps}
\affiliation{The Thacher School, 5025 Thacher Rd., Ojai, CA 93023, USA}

\author{Gazi Rahman}
\affiliation{Wesleyan University, 45 Wyllys Avenue WesBox 92055, Middletown, CT 06459, USA}

\author{Teddy Rosenbaum}
\affiliation{The Thacher School, 5025 Thacher Rd., Ojai, CA 93023, USA}

\author{Holland Stacey}
\affiliation{Stanford University, 450 Serra Mall, Stanford, CA 94305 USA}

\author[0000-0003-4454-8652]{Piper Stacey}
\affiliation{Dartmouth College, 78 College Street, Hanover, NH 03755, USA}

\author{Hadrien Tang}
\affiliation{The Thacher School, 5025 Thacher Rd., Ojai, CA 93023, USA}

\author{Asher Wood}
\affiliation{The Thacher School, 5025 Thacher Rd., Ojai, CA 93023, USA}

\author{Alejandro Wilcox}
\affiliation{The Thacher School, 5025 Thacher Rd., Ojai, CA 93023, USA}

\author[0000-0002-0184-8754]{Christopher R. Vyhnal}
\affiliation{The Thacher School, 5025 Thacher Rd., Ojai, CA 93023, USA}

\author[0000-0001-7823-2627]{Grace Yang}
\affiliation{The Thacher School, 5025 Thacher Rd., Ojai, CA 93023, USA}

\author{Jennifer Yim} 
\affiliation{The Thacher School, 5025 Thacher Rd., Ojai, CA 93023, USA}
 
\author{Yao Yin}
\affiliation{The Thacher School, 5025 Thacher Rd., Ojai, CA 93023, USA}

\author{Jack Zhang} 
\affiliation{The Thacher School, 5025 Thacher Rd., Ojai, CA 93023, USA}

\author[0000-0002-2445-5275]{Ryan J. Foley}
\affiliation{Department of Astronomy and Astrophysics, University of California, Santa Cruz, CA 95064 USA}

\author{Paul Gardner}
\affiliation{Observatory Systems, 8201 164th Ave NE \#200. Redmond, WA 98052 USA}

\author{Greg Stafford}
\affiliation{Observatory Automation Solutions, 2144 W Frostwood Ln. Tucson, AZ 85745 USA, grstafford@yahoo.com}

\author{David Rowe}
\affiliation{PlaneWave Instruments Inc., 1375 North Main St., Bldg. \#1 Adrian, MI 49221 USA}

\author{Kevin Ivarsen}
\affiliation{PlaneWave Instruments Inc., 1375 North Main St., Bldg. \#1 Adrian, MI 49221 USA}

\author{Richard Hedrick}
\affiliation{PlaneWave Instruments Inc., 1375 North Main St., Bldg. \#1 Adrian, MI 49221 USA}



\begin{abstract}
Located on the campus of the Thacher School in Southern California, the Thacher Observatory has a legacy of astronomy research and education that dates back to the late 1950’s. In 2016, the observatory was fully renovated with upgrades including a new 0.7\,m telescope, a research grade camera, and a slit dome with full automation capabilities. The low-elevation site is bordered by the Los Padres National Forest and therefore affords dark to very dark skies allowing for accurate and precise photometric observations. We present a characterization of the site including sky brightness, weather, and seeing, and we demonstrate the on-sky performance of the facility. Our primary research programs are based around our multi-band photometric capabilities and include photometric monitoring of variable sources, a nearby supernova search and followup program, a quick response transient followup effort, and exoplanet and eclipsing binary light curves. Select results from these programs are included in this work which highlight the broad range of science available to an automated observatory with a moderately sized telescope.
\end{abstract}

\keywords{telescopes --- techniques: photometric --- site testing --- supernovae: individual (SN2020hvf) --- stars: individual (Qatar 1)}


\section{Background}
\label{sec:background}
The Thacher School is a college preparatory boarding school located in Ojai, California and founded in 1889 by Sherman Day Thacher. A personal connection between Sherman Thacher and George Ellery Hale---the Director of the Mount Wilson Solar Observatory at the time---led to a rich history in astronomy at the School that is outlined by \cite{Vyhnal2018}. In 1999, the observatory fell out of use and a renovation plan was begun in earnest starting in late 2014. 

Funds for a full renovation of the observatory were secured in April of 2016, and the renovated observatory was fully functional by December of that same year. The major budget items included an upgraded 16.5-foot Ash Dome, custom dome automation hardware provided by Observatory Automation Solutions, a PlaneWave CDK-700 telescope, an Andor iKON-L 936 camera, two computers, and a pier expansion to accommodate the altitude-azimuth mount of the telescope. The design and hardware choices were informed in large part by the work of \cite{Swift2015a} in the development of the MINERVA array which has been in operation atop Mt. Hopkins in southern Arizona since 2015 \cite{Wilson2019}.

The goals of the observatory renovation were an extension of the original vision that led to the construction of the observatory on campus in 1965: to inform and inspire students about careers in science and technology \citep{Vyhnal2018}. Now, more than a half century later, we have developed an educational curriculum aimed at building practical science knowledge, skills, and experience. The educational and research goals of the Thacher Observatory go hand in hand as the pedagogical strategy is to engage students in meaningful and relevant astronomical research as a means for learning physics, astronomy and programming but also to provide a motivating force that encourages them to strive for excellence in quantitative disciplines, build a diverse and practical tool set for collaboration, and hone presentation skills.

The purpose of this paper is to outline the salient characteristics and capabilities of the renovated Thacher Observatory as a reference for our various research efforts and also to share the design and performance with the greater astronomical community, including institutions that may be interested in building an automated, small-telescope observatory for both research and education. We detail the key observatory hardware and software in \S\ref{sec:components}, discuss the characteristics of the site in \S\ref{sec:site}, and present recent results from our ongoing science programs in \S\ref{sec:science} as a proof of concept for our system. 

\section{Key Observatory Components}
\label{sec:components}
\subsection{The CDK-700 by PlaneWave}
\label{sec:telescopes}
\begin{figure}
\centering\includegraphics[width=\columnwidth]{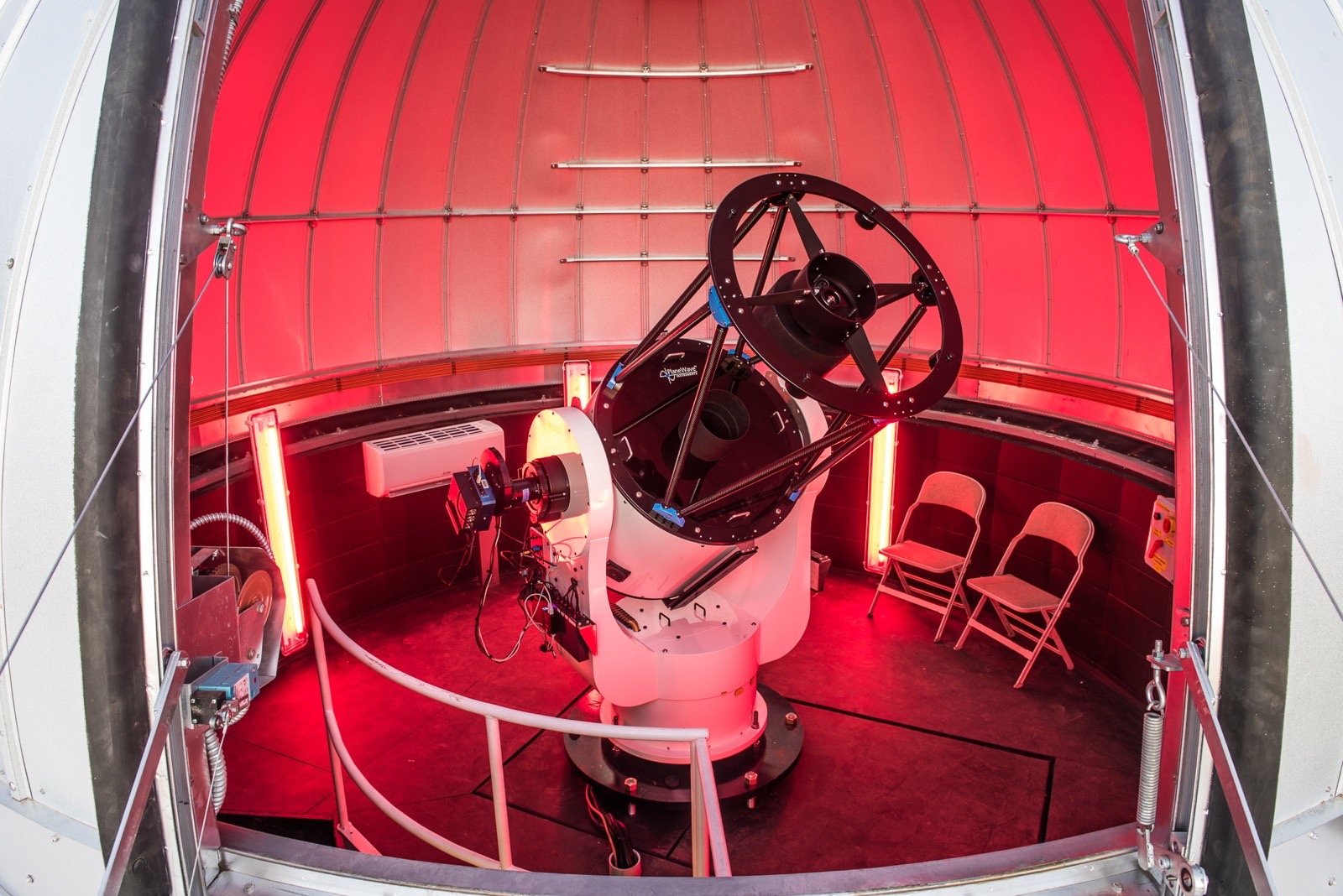}
\caption{\label{fig:cdk700}The Thacher Observatory PlaneWave CDK-700, shown inside the new $16.5^\prime$ Ash Dome.} 
\end{figure}

The PlaneWave CDK-700 is a 0.7 meter, altitude/azimuth mounted telescope system\footnote{\url{http://planewave.com/products-page/cdk700}} \citep{Hedrick2010}. Extensive testing was performed on this system through the design review phase of the MINERVA project \citep{Swift2015a}, and Thacher's telescope is essentially identical to the MINERVA telescopes. The Thacher telescope also has a mechanical primary mirror cover that is controlled remotely to protect the primary mirror from dust and debris when not in operation. Full specifications for the CDK-700 are provided by \citet[][their Table 1]{Swift2015a}. Figure~\ref{fig:cdk700} shows an image of the the Thacher Observatory telescope soon after it was installed in the renovated dome. 

The telescope pointing is controlled by two direct--drive electromagnetic motors with encoder resolution of 81\,mas resulting in a pointing accuracy of $10^{\prime\prime}$ RMS, a pointing precision of $2^{\prime\prime}$, and a tracking accuracy of $1^{\prime\prime}$ over a three-minute period. These numbers are typical but dependent on the quality of the mount model. Our mount model consists of 90 pointings distributed over the full azimuth and altitude range and has remained accurate for 3 years. The slew rate is $10^\circ$\,sec$^{-1}$ which keeps slew times between any two points on the sky to less than $20$ seconds, including settling.

The focus mechanism and image de-rotator are combined into a single, motor-controlled unit. The rotator moves at $1.6^\circ$/s, has a settling time of about 12 seconds, and has a $367^\circ$ full range of motion. When maintaining a constant position angle, the rotator can constitute a large fraction of the overhead moving time from target to target. It is standard practice to take images at a position angle of $0^\circ$ for southern sources and $180^\circ$ for northern sources thereby limiting rotator slews. Cooling fans and temperature sensors are used to keep the primary mirror in thermal equilibrium, and the control software is built to automatically correct for perturbations such as wind gusts.

\subsection{16.5-foot Ash Dome with Custom Electronics}
\label{sec:dome}
The original dome on the observatory was an early generation of Ash domes. However, the dimension of Ash domes has remained relatively constant over the years and we were therefore able to mount a modern Ash dome on the same masonry with minimal modification. Some of the new features essential for our operation are a larger, $54^{\prime\prime}$ slit and a slip ring for distributing power to the slit motor. Our dome employs an Observatory Automation Solutions' Dome Control System, which encompasses custom dome circuitry and integrated with the MaxDome II Observatory Dome Control System board set by Diffraction Limited\footnote{\url{https://diffractionlimited.com/}}.

The dome rotates at a speed of 2.8$^\circ$/s and has a two-second delay and short ramp up for smooth starts and stops. This motion is significantly slower than our telescope and can be the largest observing overhead time if the image rotator does not have a significant slew. Although the dome is computer controlled, there are also terminals on the MaxDome II board that allow the direct communication between a weather system and the dome. This allows the weather system to directly initiate a dome slit closure in order to protect the telescope, instrument, and peripheral equipment from damage by the outside elements. In addition, should the main computer crash or disconnect from the dome controller, the dome slit will close automatically after 10 minutes.

\subsection{Cameras and Filters}
\label{sec:camera}

\begin{figure}
\includegraphics[width=\columnwidth]{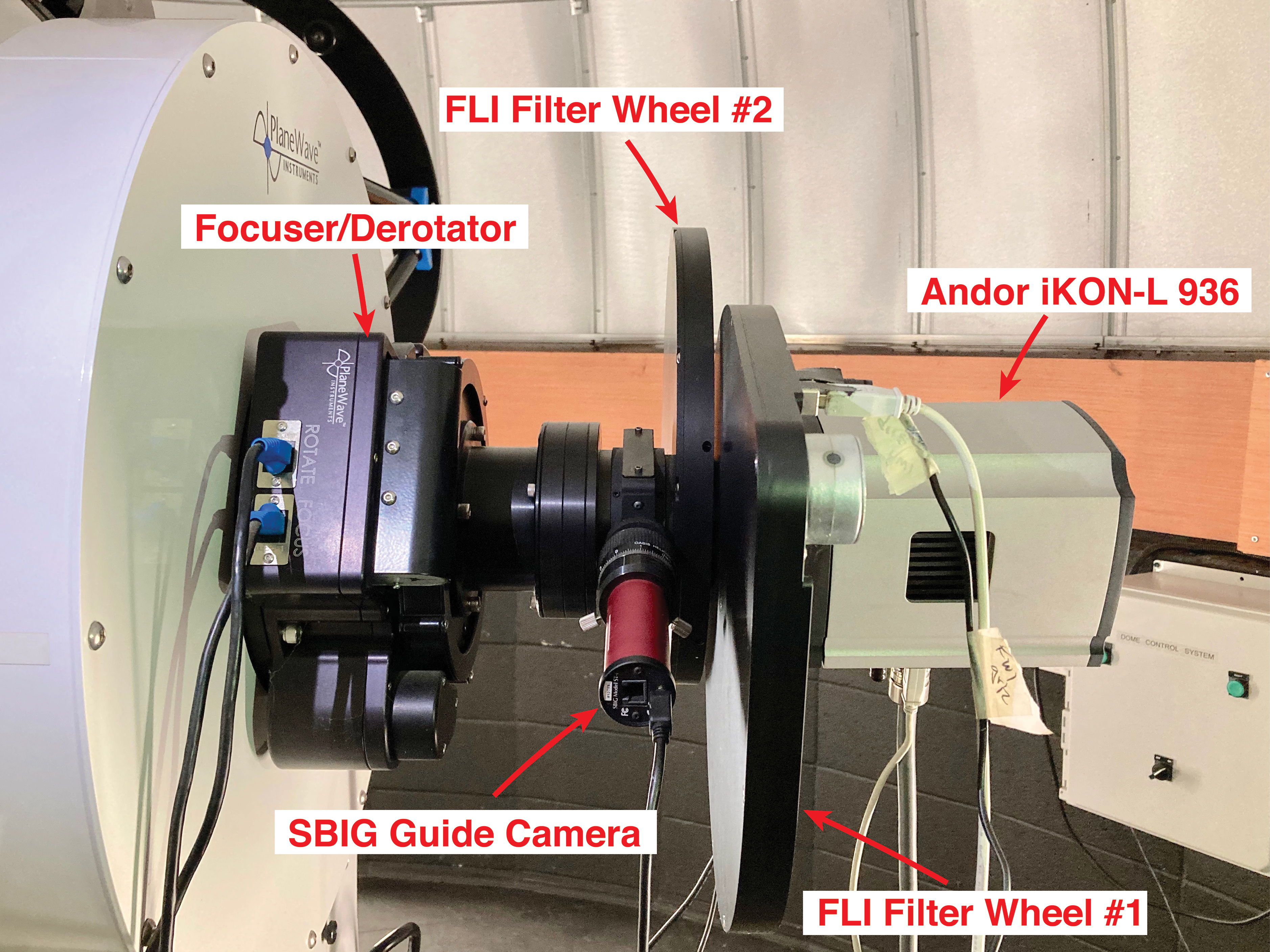}
\caption{\label{fig:camera}The optics chain for the science port of the Thacher Observatory CDK 700 with major components along the light path to the camera labeled.} 
\end{figure}

\begin{figure}[!t]
\includegraphics[width=\columnwidth]{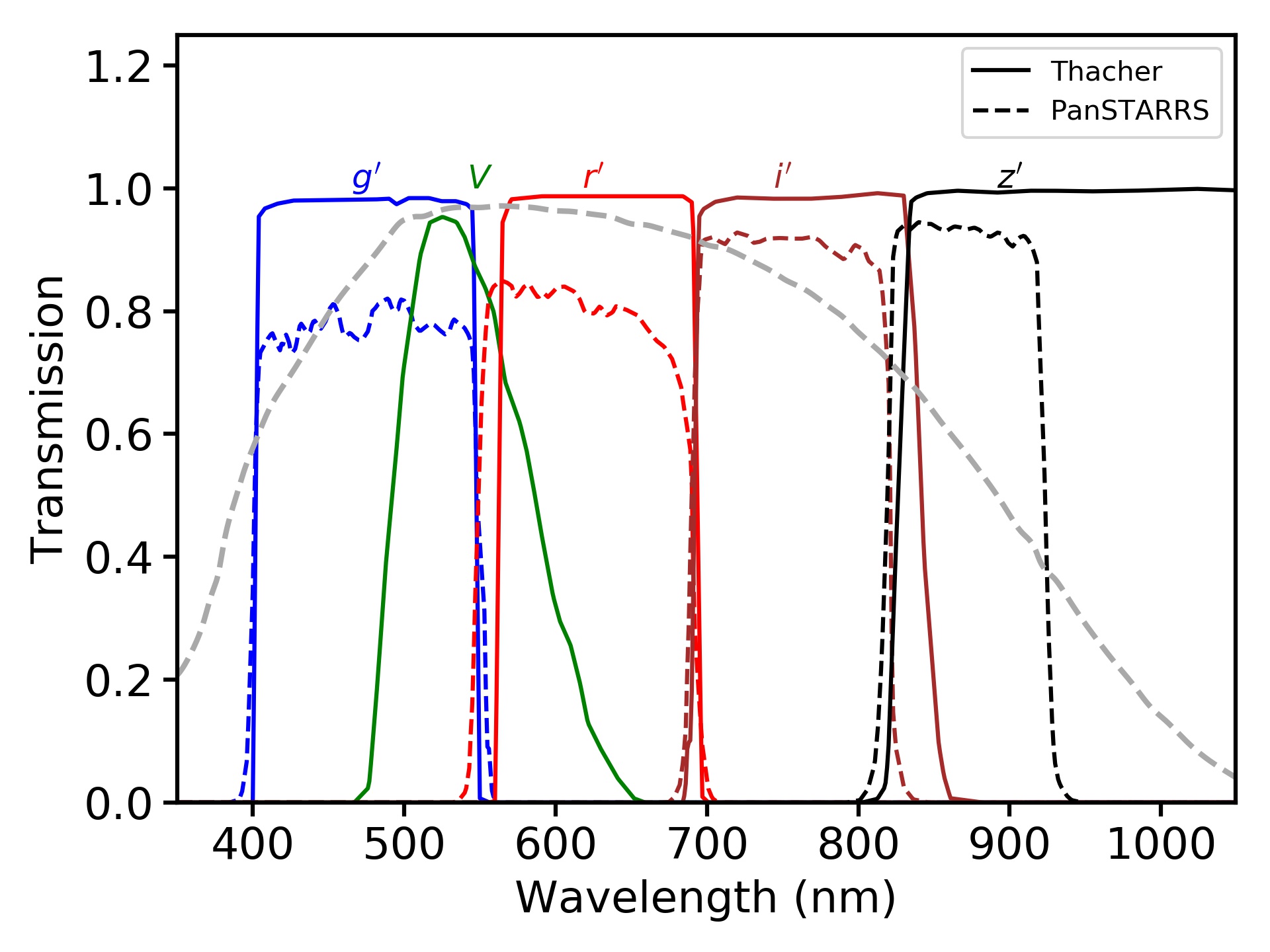}
\caption{\label{fig:filters} Transmission functions for the wide-band science filters at the Thacher Observatory as well as the quantum efficiency of our science camera. The Pan-STARRS filter band passes are shown for reference.} 
\end{figure}

Figure~\ref{fig:camera} shows an image of one of the Nasmyth ports on our CDK-700 equipped with several components, including our filter wheels, an SBIG ST-i guide camera, and our science camera. Dual, 10-slot filter wheels from Finger Lakes Instruments which fit 50\,mm square filters are mounted back to back with at least one empty slot per filter wheel allowing access to as many as 18 different filters. All our filters were purchased online from Astrodon\footnote{\url{https://astrodon.com/}}. One filter wheel is equipped with our science filters: a Sloan second generation set $g^\prime$, $r^\prime$, $i^\prime$, and $z^\prime$, a Johnson-Cousins $V$-band filter, and a set of 5\,nm narrow band filters for H$\alpha$, [S{\sc ii}], and [O{\sc iii}]. The filter transmission functions as well as the quantum efficiency of our camera are plotted in Figure~\ref{fig:filters}. The second filter wheel contains our astrophotography set, L, R, G, and B as well as two diffusers from RPC Photonics\footnote{\url{https://www.rpcphotonics.com/}}, EDC-0.25 which provides a $0.25^\circ$ diffusion angle and EDC-0.5, which provides a $0.5^\circ$ diffusion angle. The second Nasmyth port on our telescope is dedicated to eyepiece observing.

Our imaging camera is an Andor iKON-L 936\footnote{\url{http://www.andor.com/scientific-cameras/ikon-ccd-camera-series/ikon-l-936}} equipped with a back illuminated sensor with wide band (BV) coating and $2048\times2048$ square 13.5\,\micron\ pixels for a total chip size of 27.6\,mm corresponding to a $20.8^\prime$ field on our telescope. The camera has a circular, integrated shutter system. The manufacturer stated shutter speed is 15\,ms from half open to half closed, which limits the exposure time to be greater than a few seconds to ensure that the ratio of flux in the outer half of the chip is less than 1\% different than the flux in the inner half. Exposure times of $\gtrsim 15$ seconds should be used to ensure millimagnitude level accuracy. Also, the circular mechanism that holds the shutter vignettes our sensor at the corners blocking approximately 2\% of the sensor at the corners of the chip. This blockage at the extreme corners of the chip do not impact science observations. 

\subsection{Weather Monitoring Equipment}
\label{sec:weathereq}
The Thacher Observatory is designed to be a fully automated facility. Therefore it is critical that the weather is monitored reliably, and that the dome can be shut if weather conditions threaten the observatory equipment. 

For long-term monitoring of the weather we use a Davis Vantage Pro2 Plus weather station located east of the storage shed approximately 200 feet northeast of the observatory. We have data from this instrument dating to back to April, 2015, which includes 27 columns of information.

Although the Davis weather station provides comprehensive information on weather conditions, it cannot detect the presence of clouds--a critical task for keeping the observatory equipment safe. For this task we have used two instruments: the Boltwood Cloud Sensor II by Diffraction Limited, and, since the Spring of 2020, the AAG CloudWatcher by Lun\'{a}tico. Both of these units detect cloud coverage by measuring a sky temperature and have an API that allows communication with observatory control software. In addition, both units have the ability to send a panic signal along a dedicated wire to instigate a dome closure. Cloud sensor data exist from November 2016, but the interpretation of the data is not always straightforward as the degradation of the sensors and changing atmospheric conditions affect the temperature threshold between cloudy and clear. 

In addition to this instrumentation, a Santa Barbara Instruments Group (SBIG) Seeing Monitor was purchased for the observatory in 2015 and has been used since to track the expected image quality at zenith. The unit has a fixed mount oriented toward the north celestial pole. It takes fast images of Polaris and monitors the brightness and centroid fluctuations to calculate an estimate for the expected image quality at zenith. Because this instrument is an older model that is no longer available and has limited technical support, observatory data are sparse after 2018.

\subsection{Observatory Control and Other Components}
\label{sec:control}
The Thacher Observatory telescope is controlled by the PlaneWave Interface 2\footnote{\url{https://planewave.com/software/\#section-pwi2}} (PWI2) software, and the Andor iKON-L camera and Fingerlakes filter wheel is controlled by Maxim DL, which interfaces with PWI2 for tasks such as auto-focusing. We also employ Planewave Shutter Control for our primary mirror cover. All this software communicates via standard ASCOM protocols, which is essential for the automation software ACP Expert developed by DC3-Dreams. ACP Expert employs a dispatch scheduler that reads and optimizes observations from a queue populated with constraints and priorities that is updated at the conclusion of each observation. 

Many of our observations require accurate and precise timing. Therefore the Thacher Observatory has a local GPS server---NTP100-GPS Time Server from Masterclock---that is synced to our Windows 10 system. Since November 1, 2018, we have been using this server to update our system time.

\section{Observatory Site}
\label{sec:site}

The observatory in its original form was erected on the Thacher campus in April of 1965 using funds obtained by George Abell of the University of California, Los Angeles \citep{Vyhnal2018}. While the location was undoubtedly chosen based on the successful Summer Science Program (SSP) that had been started six years earlier under the direction of Thacher's Headmaster at the time, Newton Chase, the details of the site selection process are unclear. Therefore we have compiled our own, recent data to gain a broader understanding of the site.

\begin{deluxetable*}{@{\extracolsep{0.1in}}rl}
\tablenum{1}
\tablecaption{Thacher Observatory Overview \label{tab:observatory}}
\tablehead{\multicolumn{2}{c}{Location}}
\startdata
City\dotfill                         & Ojai, CA  \\
Latitude\dotfill                     & $38^\prime$ $00.5^{\prime\prime}$ N \\
Longitude\dotfill                    & $119^\circ$ $10^\prime$ $38.5^{\prime\prime}$ W\\
\hline
\multicolumn{2}{c}{Weather} \\
\hline
Clear Nights Per Year\tablenotemark{a}\dotfill                   & $\sim 270$\\
Average Yearly Rainfall\tablenotemark{b}\dotfill                 & $\sim 19^{\prime\prime}$\\
\hline 
\multicolumn{2}{c}{Sky Brightness (Moonless)} \\
\hline 
$V$-band\dotfill    & 21.7 mag/arcsec$^{2}$ \\
$g$-band\dotfill    & 22.1 mag/arcsec$^{2}$ \\
$r$-band\dotfill    & 21.9 mag/arcsec$^{2}$ \\
$i$-band\dotfill    & 21.3 mag/arcsec$^{2}$ \\
$z$-band\dotfill    & 20.0 mag/arcsec$^{2}$ \\
\hline 
\multicolumn{2}{c}{Atmospheric Conditions} \\
\hline 
Median Seeing\dotfill         & $3.0^{\prime\prime}$ \\
$V$-band extinction\dotfill   & 0.14 mag/airmass \\
$g$-band extinction\dotfill   & 0.19 mag/airmass \\
$r$-band extinction\dotfill   & 0.08 mag/airmass \\
$i$-band extinction\dotfill   & 0.05 mag/airmass \\
$z$-band extinction\dotfill   & 0.02 mag/airmass \\
\hline 
\multicolumn{2}{c}{Zero Point Magnitudes} \\
\hline 
$V$-band\dotfill   & 21.73 mag \\
$g$-band\dotfill   & 22.08 mag \\
$r$-band\dotfill   & 21.91 mag \\
$i$-band\dotfill   & 21.31 mag\\
$z$-band\dotfill   & 20.04 mag\\
\hline
\multicolumn{2}{c}{Telescope} \\
\hline 
Design\dotfill                          & Corrected Dahl-Kirkham \\
Aperture\dotfill                        & 0.7\,m \\
Focal Ratio\dotfill                     & f/6.5 \\
Central Obscuration\dotfill             & 47\% \\
Pointing Accuracy\dotfill               & $<10^{\prime\prime}$ \\
Tracking Precision\dotfill              & $< 1.5^{\prime\prime}$ over 5\,min 
\enddata
\tablenotetext{a}{Approximate number from observatory weather station and historical records}
\tablenotetext{b}{Actual rainfall year-to-year can vary greatly from $\sim 7$ inches to 40+ inches. The average is influenced strongly by a small number of very wet years.}
\end{deluxetable*}

\begin{figure}
\includegraphics[width=\columnwidth]{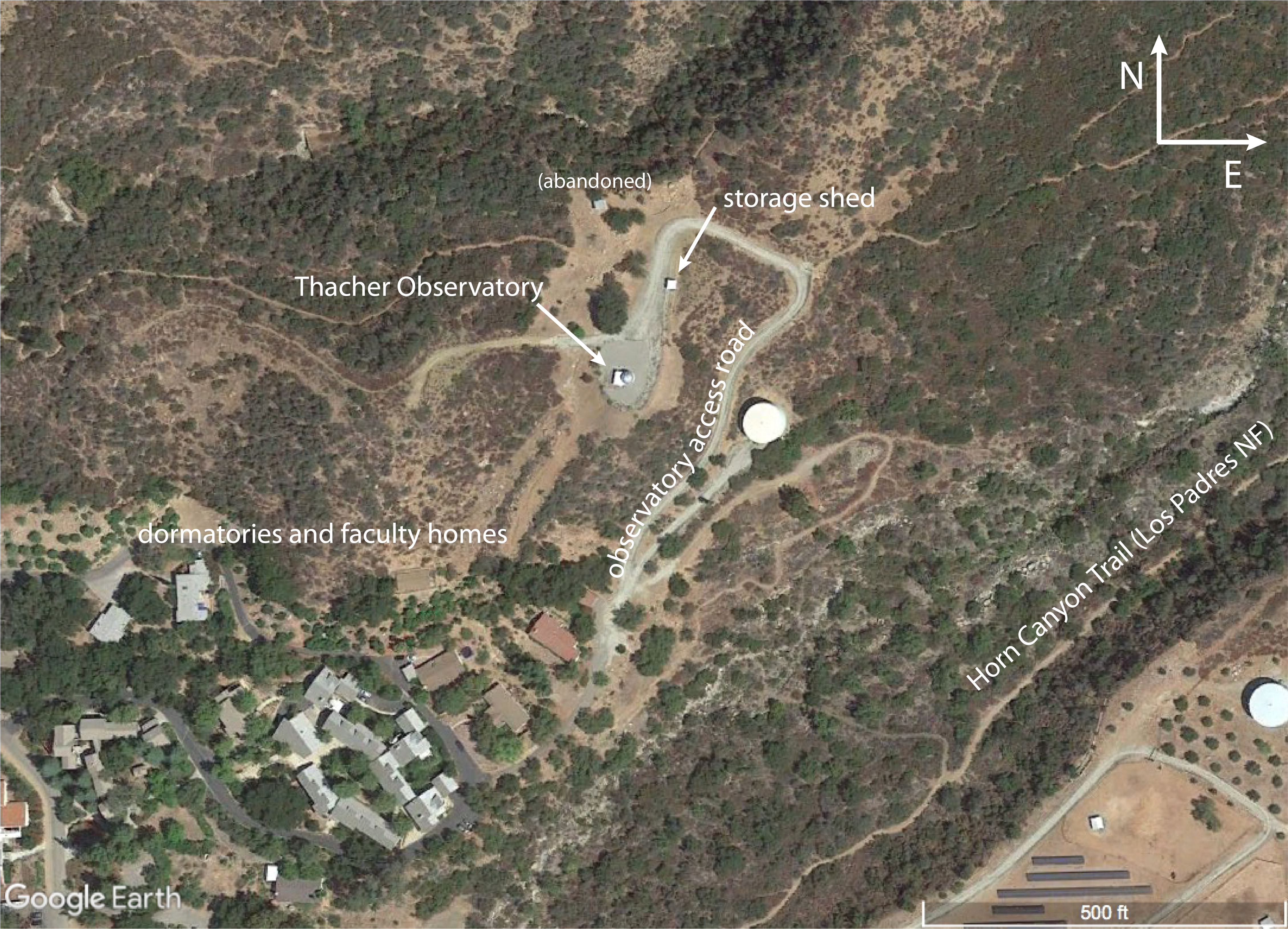}
\caption{\label{fig:site} Annotated image of the Thacher Observatory and surroundings. The observatory is located at ($\phi$, $\lambda$) = $34^\circ$ $38^\prime$ $00.5^{\prime\prime}$ N, $119^\circ$ $10^\prime$ $38.5^{\prime\prime}$ W at an elevation of 495\,m (1620 ft) and is seen near the center of the image. } 
\end{figure}

\vspace{-24pt}
The observatory is located at ($\phi$, $\lambda$) = $34^\circ$ $38^\prime$ $00.5^{\prime\prime}$ N, $119^\circ$ $10^\prime$ $38.5^{\prime\prime}$ W at an elevation of 495\,m (1620 ft). Figure~\ref{fig:site} shows a Google Earth\footnote{\url{http://earth.google.com/web/}} image of the site which lies in the northeastern corner of the developed part of the Thacher campus and borders the Los Padres National Forest.

The local horizon is lowest in the West where the sky is visible down to a few degrees in altitude, and it is highest to the North where the local mountains obstruct the sky up to approximately $18^\circ$. The local horizon line, on average, is at about $7^\circ$ above the true horizon.

\subsection{Weather}
We use our Boltwood CloudSensor II data to estimate the average fraction of nighttime when observing is possible. This is calculated using the sky temperatures during times when the darkness of the sky is below the astronomical twilight threshold. Both of these quantities are measured by the sensor while the thresholds have been set manually. The largest source of uncertainty in determining this number originates from the fact that the transparency of the thermopile window on the Boltwood CloudSensor II degrades with time, and therefore the sky temperature threshold between conditions favorable and unfavorable for observations changes. However, these changes are relatively slow. Our best estimate for the fraction of nighttime when conditions are favorable for observations is 80\%. This would translate to approximately 290 nights per year when observing is possible as compared to $\sim 270$ nights per year that are considered clear. It should be noted that the time over which these data were taken (approximately 4 years from 2016 through 2019) were drought conditions.

\subsection{Sky Brightness}
There were no sky brightness measurements known when plans to renovate the observatory were being made. However, the relative darkness of the night sky in the Ojai valley is obvious and some detail in the Milky Way can be seen with the naked eye on moonless nights. The town borders the Los Padres National Forest where light pollution is minimal, and a dark-sky ordinance has been in place in the city of Ojai since November, 2018.

To quantify the sky brightness at the site of the Thacher Observatory, which sits immediately adjacent to the Los Padres National Forest on the outskirts of Ojai, we used a calibrated photometer on loan from the Palomar Observatory. The photometer was built by D. McKenna and used to measure the sky brightness at the Palomar Observatory. It has a $5^\circ$ field of view, and the V-band surface brightness in magnitudes per square arcsecond can be derived from the raw count rate using:
\begin{equation}
S_V = -2.5\log_{10}(C-D)+18.865
\label{eq:skybrightness}
\end{equation}
where $C$ is the raw count rate as read by the unit, and $D$ is the dark count rate.
(D. McKenna, 2015; personal communication).

On the evening of April 12, 2015 at 11:30 pm local time, we obtained simultaneous images with our all-sky camera and measurements with the photometer pointed at zenith. It was a moonless night with the Galactic anti-center low on the western horizon at the time. Multiple readings with the photometer averaged to a count rate of 0.1512 which translates to a sky brightness of 21.029 magnitudes per square arcsecond. 

\begin{figure}
    \centering
    \includegraphics[width=\columnwidth]{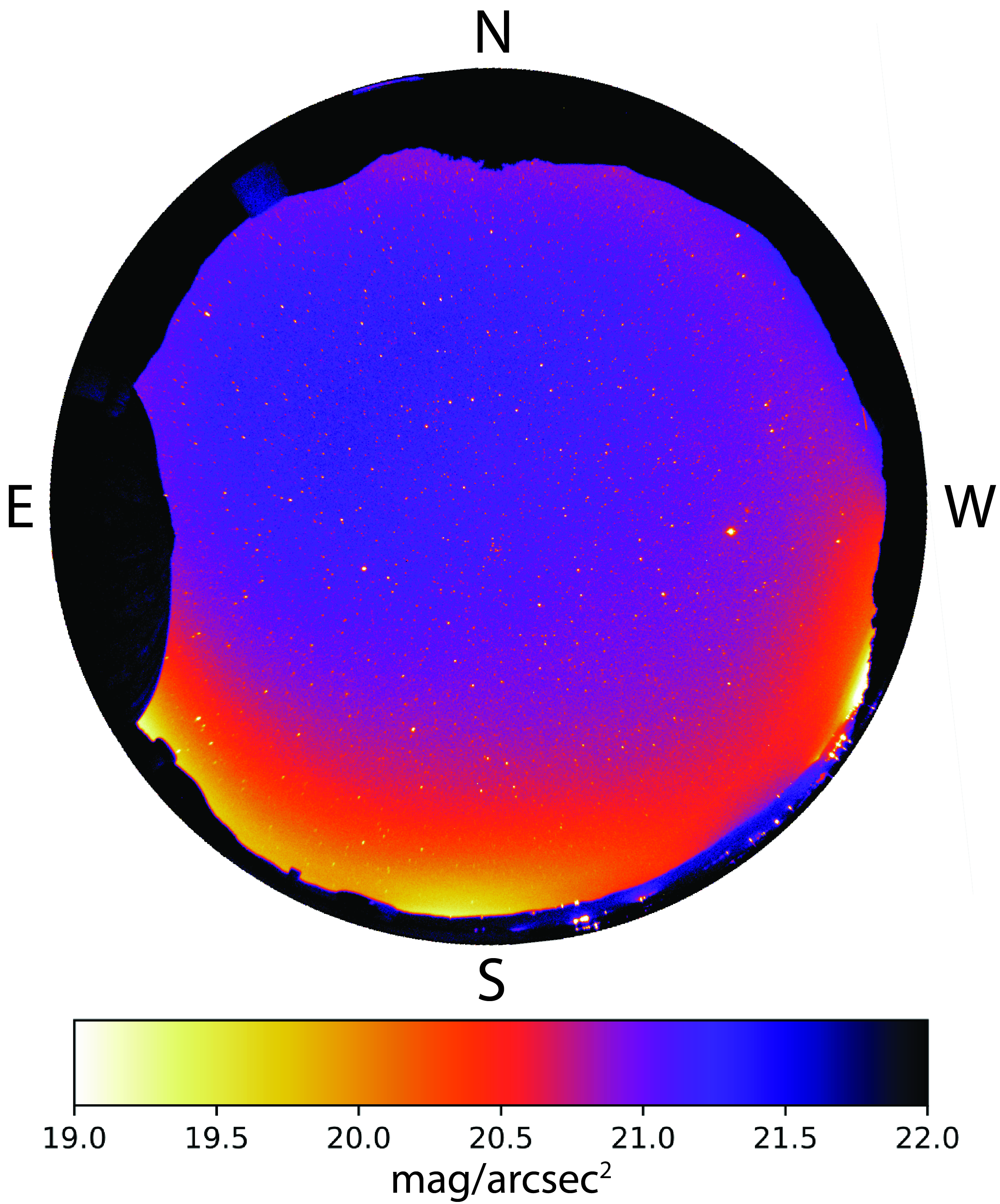}
    \caption{All-sky image taken from the Thacher Observatory calibrated for V-band surface brightness near zenith. }
    \label{fig:skybrightness}
\end{figure}

An all-sky image taken at the same time as the photometer readings was calibrated using dark frames taken earlier that evening, and the center section of the image was plate solved using \texttt{astrometry.net} \citep{Lang2010}. Zenith was located using the observatory coordinates and local sidereal time at the time of the image. A $2.5^\circ$ aperture was constructed around zenith and the average counts were calculated within the aperture. These counts were then used to scale the image to the photometer reading. The result is shown in Figure~\ref{fig:skybrightness}. The artificial glow from Los Angeles is hidden mostly behind the dome, while the horizon line to the south and southeast is dominated by city lights from Ventura. The glow to the West is from Santa Barbara. The ridge of brightness toward the horizon in the northwest is actually the Galactic plane and the bright source in the west is Jupiter. It can be seen that the darkest section of the sky is actually north of zenith, toward the National Forest. 

We have since used science images to validate this measurement and also to estimate the sky brightness in our other science bands. We found that the sky brightness can vary significantly---by a magnitude or more---even for moonless, photometric nights. Further study is needed to understand this variability.

We also estimate the sky brightness in each of our wide photometric bands using on-sky measurements obtained during a photometric night in May 2020 while the moon was below $-18^\circ$ elevation. The sky was sampled at 5 randomly selected elevations between 15 and 85 degrees at each of 24 azimuth angles separated by 15 degree increments. Images taken above an airmass of 1.2 and more than 15 degrees away from the galactic plane were calibrated and a photometric zero point was found using Pan-STARRS sources with error and variation of less than 1\%. This zero point was applied to the median sky counts and then corrected for our $0.608^{\prime\prime}$ square pixels to derive a magnitude per square arcsecond in each of our science bands. We used \cite{Jester2005} to convert the Pan-STARRS magnitudes to $V$-band magnitudes for our $V$-band observations. The results are presented in Table~\ref{tab:observatory} with errors expected to be at the 10\% level.

\subsection{Image Quality and Seeing}
Results from our SBIG seeing monitor have been presented at past meetings of the American Astronomical Society \citep{ONeill2016,ONeill2017}. The median and mode of the expected FWHM distributions were found to be $3.5^{\prime\prime}$ and $2.7^{\prime\prime}$, respectively. Given the limited nature of those data and more than four years of science data acquired at the observatory, we performed an independent check on the image quality at the observatory by analyzing science data. 

We have been monitoring Boyajian's Star \citep{Boyajian2012,Boyajian2018} since May 2017, and our images of this field are ideal for this analysis since the field is at a Galactic latitude of $6.6^\circ$ and contains many stars, it reaches an altitude of $80^\circ$ (airmass $=1.02$) from the vantage point of our observatory, and each nightly set of observations of Boyajian's star is preceded by an autofocus in the $r^\prime$-band. 

\begin{figure}
    \centering
    \includegraphics[width=1.1\columnwidth]{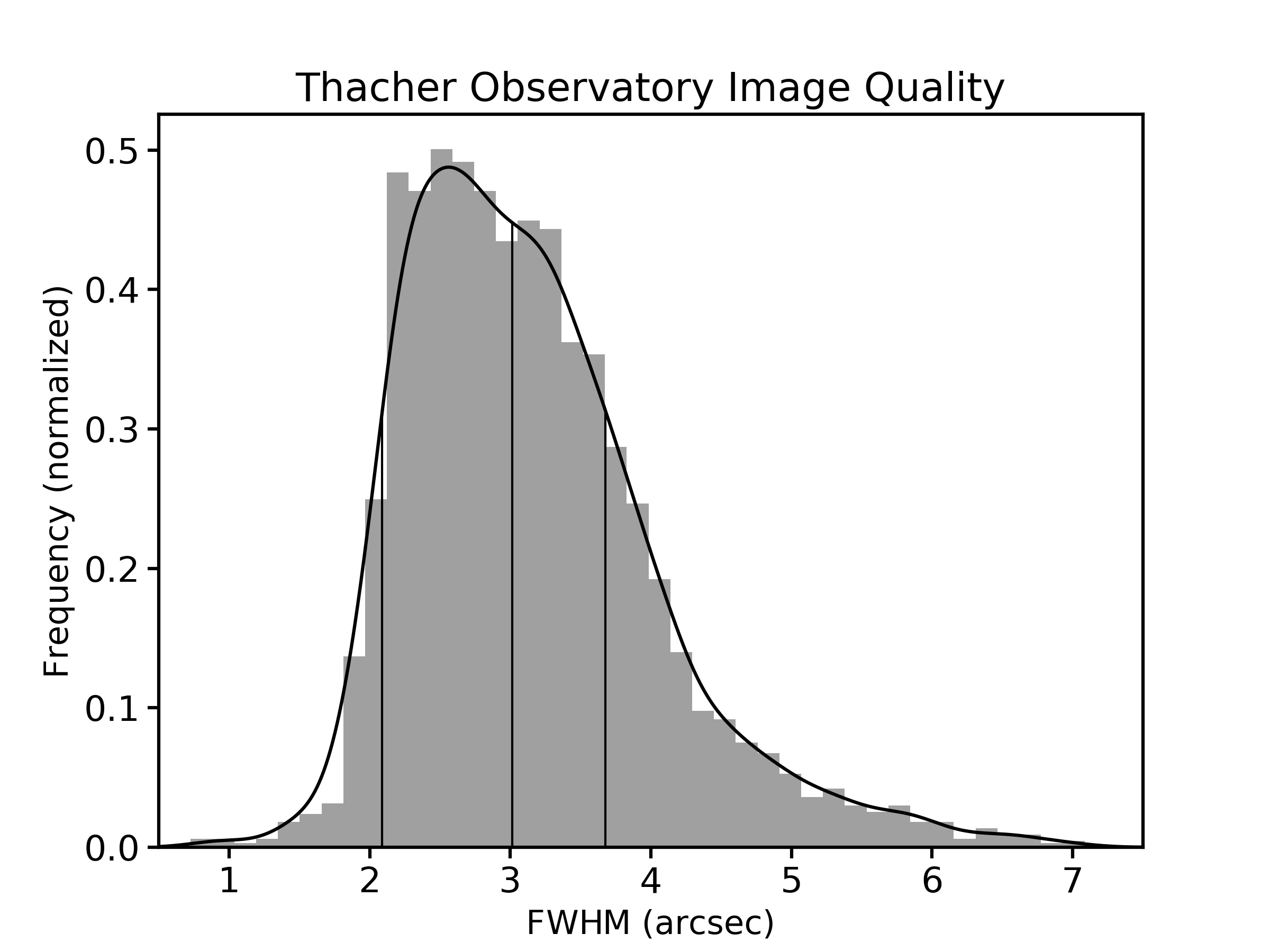}
    \caption{Histogram and kernel density estimation of point source FWHM values for 4290 images of the field around Boyajian's Star taken above an airmass of 1.1. The mode of the distribution is $2.6^{\prime\prime}$, and the median value of $3.0^{\prime\prime}$ and shortest 1-$\sigma$ interval from $2.1^{\prime\prime}$ to $3.7^{\prime\prime}$ are shown.}
    \label{fig:imagequality}
\end{figure}

To estimate the image quality we calibrate and analyze 4290 $r^\prime$-band images of Boyajian's Star taken at airmass $< 1.1$ (altitude $>65^\circ$) spanning more than four years. Source Extractor \citep{Bertin1996} was used on each image with detection and analysis thresholds of 10. The median FWHM of all sources detected (typically several hundred or more) was used to characterize the image. The results are summarized in Figure~\ref{fig:imagequality}. The median of the skewed distribution is $3.0^{\prime\prime}$, the mode is $2.6^{\prime\prime}$, and the shortest 1-$\sigma$ interval is from $2.1^{\prime\prime}$ to $3.7^{\prime\prime}$. 

\section{System Characterization}
\label{sec:characterization}
To both plan successful observations and to understand and correctly interpret our photometric data, we provide a full characterization of the Thacher Observatory including details about the performance of the Andor iKon-L camera, the CDK 700, and our control system. 
\vspace{24pt}

\subsection{The Andor iKON-L 936}
\label{sub:camera}
The Andor iKON-L 936 camera has 24 unique settings available which offer various advantages and drawbacks. Through experimentation, we found a setting that offers the stability, fast readout, low read noise, and low dark current that suits our goals for precision photometry. These settings are outlined in Table~\ref{tab:camera}. 

Although the iKON-L has a five stage thermoelectric cooling system capable of achieving temperatures down to $\sim -100^\circ$C with the use of glycol or distilled water, for standard observations we use a camera temperature of $-30^\circ$C which provides a nice balance of fast cooling time, low maintenance (no need for glycol or distilled water cycling), low dark current, and the ability to achieve operational temperature even on the hottest summer nights.

\begin{deluxetable*}{@{\extracolsep{0.1in}}rl}
\tablenum{2}
\tablecaption{iKON-L 936 BV Camera Specifications\label{tab:camera}}
\tablehead{\multicolumn{2}{c}{Design}}
\startdata
Size\dotfill                         & $2048\times 2048$ \\
Pixel Dimensions\dotfill             & $13.5\,\mu$m, square \\
\hline 
\multicolumn{2}{c}{Nominal Settings} \\
\hline 
A/D Rate\dotfill                         & 1\,MHz \\ 
Amplifier Gain\dotfill                   & $1\times$\\
Vertical Shift Speed\dotfill             & 38.55\,$\mu$sec \\
Vertical Clock Voltage Amplitude\dotfill & Normal \\
Output Amplifier\dotfill                 & High Sensitivity \\
Baseline Clamp\dotfill                   & On \\
Nominal Operating Temperature\dotfill            & $-30^\circ$C \\
\hline 
\multicolumn{2}{c}{Performance} \\
\hline 
Gain\dotfill                             & $3.59 \pm 0.02$\,\,e$^-$/ADU \\
Bias Level\dotfill                       & $300.0 \pm 0.5$\,ADU \\
Readnoise\dotfill                        & $11\pm2$\,e$^-$ \\
Dark Current\tablenotemark{a}\dotfill                     & 0.25\,e$^-$/s \\
Readout Time\dotfill                     & 4 seconds \\
Linearity better than\dotfill            & 0.2\% up to 28000\,ADU \\
Plate Scale\dotfill                      & $0.61^{\prime\prime}$/pixel\\
Well Depth\tablenotemark{b}\dotfill                       & 92072\,e$^{-}$ \\
\enddata
\tablenotetext{a}{Measured at nominal temperature of $-30^\circ$C}
\tablenotetext{b}{Provided by manufacturer}
\end{deluxetable*}

\vspace{-24pt}
While our camera came with full specifications provided by the manufacturer, we re-measured many of the fundamental properties of the camera with a series of bias, dark and flat frames acquired on 16 July 2020. Our flat frames for this experiment were taken with a white sheet taped taut to the front of the OTA which was then illuminated by a 10\,W incandescent bulb powered through an APC Smart 3000VA UPS. Forty bias frames and forty flat field frames were taken, and the flat field frames were taken in pairs at exposure times between 3 and 45 seconds. This allowed us to test the linearity of the CCD as well as the camera gain. In addition, twenty dark frames were taken with exposure times of 120\,s, adequate to provide an accurate reading of the dark current.

The camera bias was measured to be $300.0 \pm 0.5$ ADU, where the error signifies the 1-$\sigma$ pixel to pixel variation. To check the stability of the bias frames we compared the mean value of 1056 master bias frames made from standard observing nights spanning almost 4 years, from 14 January 2017 to 18 December 2020. The standard deviation of the median value of the bias frames was 0.19\,ADU. The read noise was calculated from the standard deviation of each individual camera pixel across our forty bias frames. The mean of the standard deviation of the individual pixels give an average read noise of $3.0\pm 0.6$\,ADU.

\begin{figure}[!hb]
    \centering
    \includegraphics[width=1.1\columnwidth]{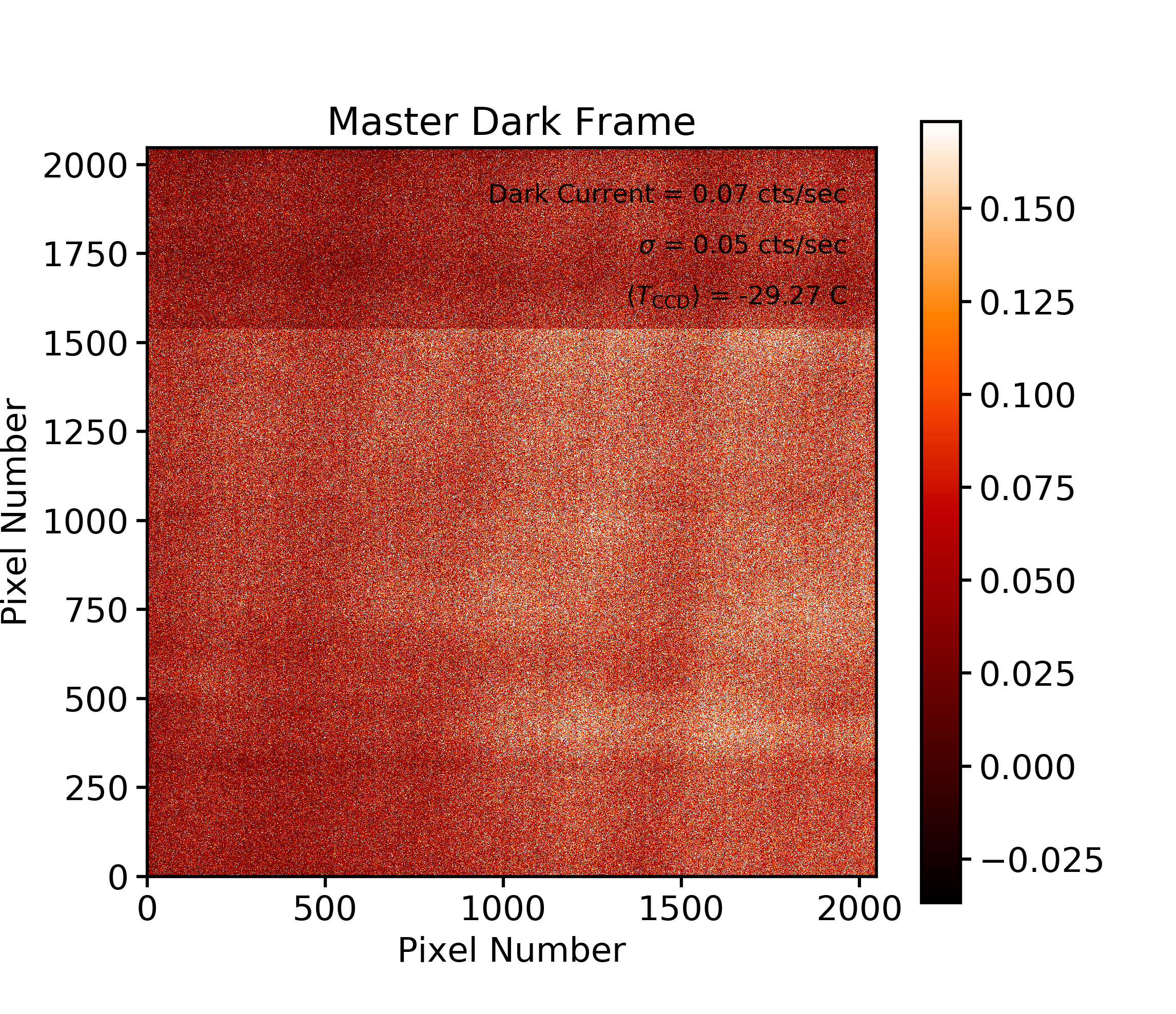}
    \caption{Master dark frame constructed from twenty 120 second exposures. The coherent features seen in the frame are due to variations in the dark current and are stable.}
    \label{fig:dark}
\end{figure}

The dark current for our camera at an operating temperature of  $-30^\circ$C was consistent with the value reported by the manufacturer, $0.07$\,ADU/s, but was also found to vary considerably across the chip with a standard deviation of $0.05$\,ADU/s. The features seen in the master dark frame in Figure~\ref{fig:dark} are stable and calibrate well even for long exposures. 

\begin{figure}
    \centering
    \includegraphics[width=1.1\columnwidth]{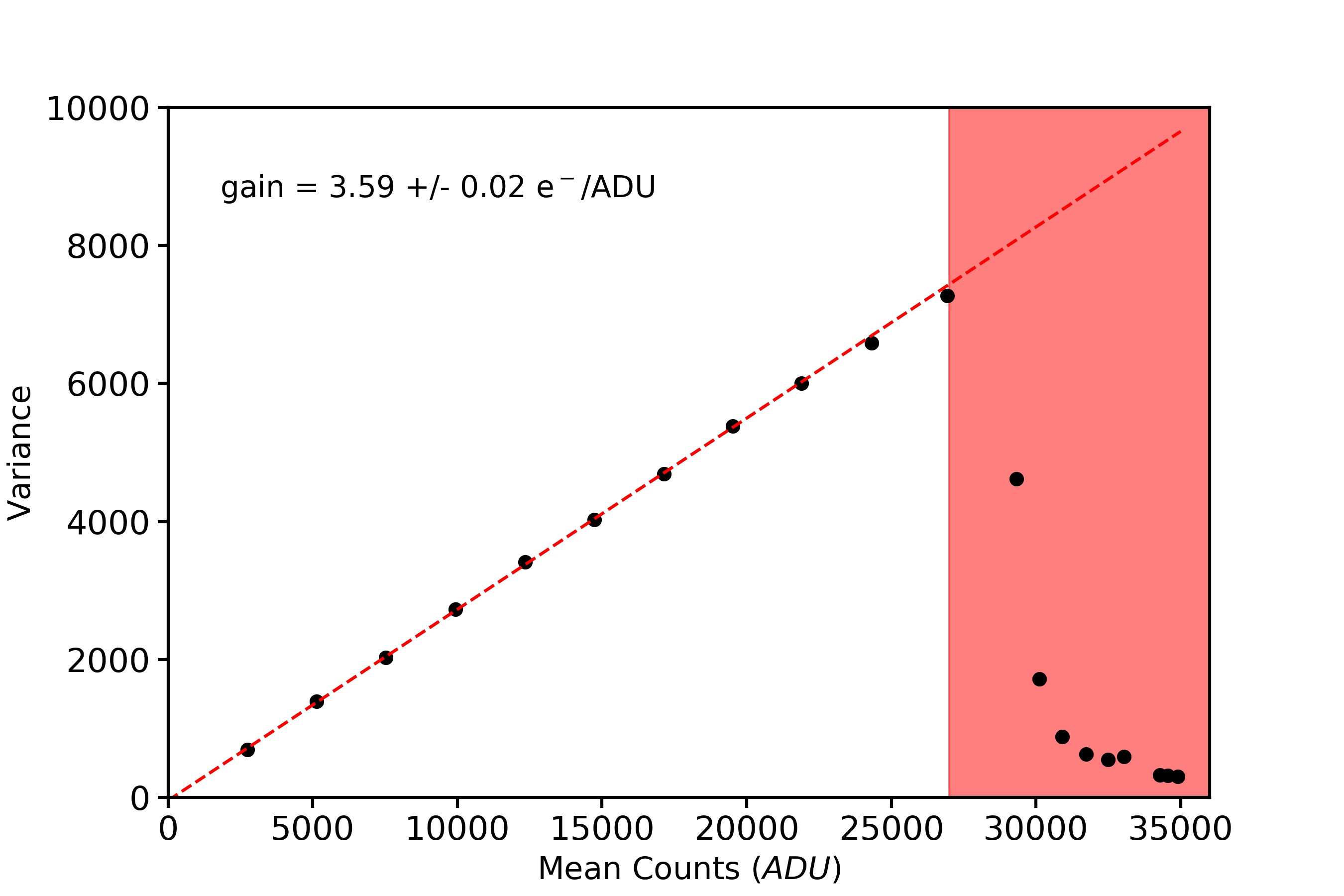}
    \caption{Variance measured in a select region of the CCD chip plotted versus the mean counts in that region. The best fit line (\texttt{red, dashed line}) was used to measure the gain of the Andor iKON-L.}
    \label{fig:gain}
\end{figure}

The linearity of our chip was tested by taking the median value of a $150 \times 150$ region of pixels near the center of the CCD for a range of exposure times spanning from 3 to 45 seconds. The response of the chip was measured to be linear at a level better than 0.08\% for mean counts $\lesssim 27000$\,ADU. Between approximately 27000 and 33000 ADU the linearity seems to show deviations up to 0.75\%. Above 32000\,ADU the chip suffers significant deviations from linearity. The region above 27000\,ADU is shaded red to reflect increased non-linearity as well a significant deviation from Poisson statistics (see Figure~\ref{fig:gain}).

\begin{figure}
    \centering
    \includegraphics[width=1.1\columnwidth]{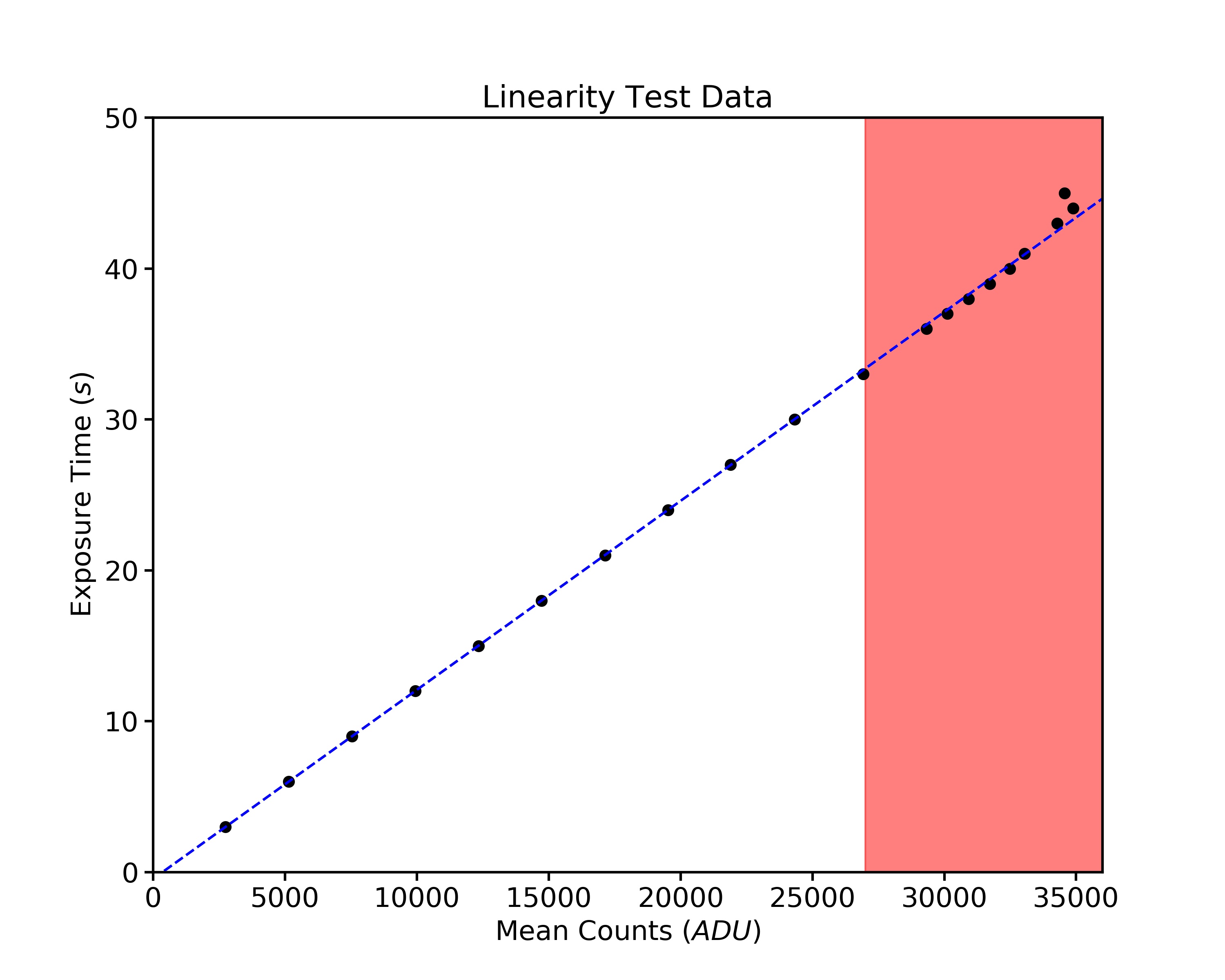}
    \caption{The exposure time of our test images plotted against the mean counts measured in a select region of the CCD chip showing the linear response of the Andor iKON-L 936.}
    \label{fig:linearity}
\end{figure}

The gain of our camera was calculated by two methods. The first was calculated using the method outlined by \citet[][\S 4.3]{Howell2006} using pairs of bias frames and flat field frames with the same exposure times for exposure times spanning from 3 to 45 seconds. The gains measured for each exposure time were consistent up to frames with mean counts beyond about 27000\,ADU. We use the mean and standard deviation of the 10 measurements below this level to find $g = 3.57\pm0.04$\,e$^{-1}$/ADU, which is consistent with the manufacturer provided value for our specific camera settings. 

We also calculated the gain using the expectation that photoelectrons follow a Poisson distribution with a variance equal to the mean, while the variance of the counts will be reduced by a factor of the gain. Using the same $150 \times 150$ region of pixels near the center of the CCD as used for the linearity test, we plot the variance versus the mean counts in each pair of frames. It can be seen in Figure~\ref{fig:gain} that the statistics of the pixels start to deviate from what is expected beyond mean counts of about 27000\,ADU. This corresponds to the mean count level where the linearity of the chip also starts to show signs of degradation. Fitting a line to the data below this mean count level results in a gain estimation of $g = 3.61\pm0.02$\,e$^{-1}$/ADU. Averaging the gain measurements obtained by these two different methods yields a final gain estimate of $g = 3.59\pm0.02$\,e$^{-1}$/ADU. Based on the statistical behavior and the measured linearity of the our CCD, we suggest that that peak counts are kept to levels below 27000\,ADU to ensure the most accurate and precise photometric measurements.

\subsection{Telescope Zero Points and Atmospheric Extinction}
\label{sec:mzp}
A common and effective way to characterize the performance of an observatory is through the zero point magnitudes and atmospheric extinction coefficients which can be measured in each photometric band of interest. The zero point magnitude is defined as the magnitude of a source that would produce one count per second if there were no atmospheric extinction. 

For observations of a field with sources of known magnitude, we can define the photometric zero point, $ZP$ to be the magnitude of a source that would produce one count per second at the given airmass and conditions by rearranging the definition of apparent magnitude
\begin{equation}
    m - m_{ref} = -2.5\log_{10}\left(\frac{S}{S_{ref}}\right)
    \label{eq:mag}
\end{equation}
By setting $S=1$, $m$ becomes the photometric zero point and the reference magnitude, $m_{ref}$ will be equal to the known magnitude of a source associated with $S_{ref}$ which is measured in ADU/s from the image.

\begin{equation}
    ZP = m_{ref} + 2.5\log(S_{ref})
    \label{eq:IM}
\end{equation}

Because the atmospheric extinction diminishes the observed flux from a source by an exponential factor and magnitudes are a logarithmic measure of brightness, the photometric zero point as a function of airmass will follow a linear trend. Extrapolating this trend to zero airmass will then yield the zero point magnitude, $m_{zp}$.

On the night of UT 9 June 2020 under photometric conditions, we sampled the entire sky in steps of $15^\circ$ in azimuth, with four 30 second observations in each of our wide photometric bands---$g^\prime$, $r^\prime$, $i^\prime$, $z^\prime$, and $V$---at each azimuth chosen randomly between $15^\circ$ and $85^\circ$ altitude resulting in 384 observations of the sky at a variety of airmasses. We used DoPHOT \citep{Schechter1993} to obtain PSF fitted fluxes and then matched targets in the Pan-STARRS catalog \citep{Flewelling2020,Magnier2020} brighter than 16th magnitude in $r$ and with error and standard deviation both less than 2\%. 
\begin{figure}
    \centering
    \includegraphics[width=1.1\columnwidth]{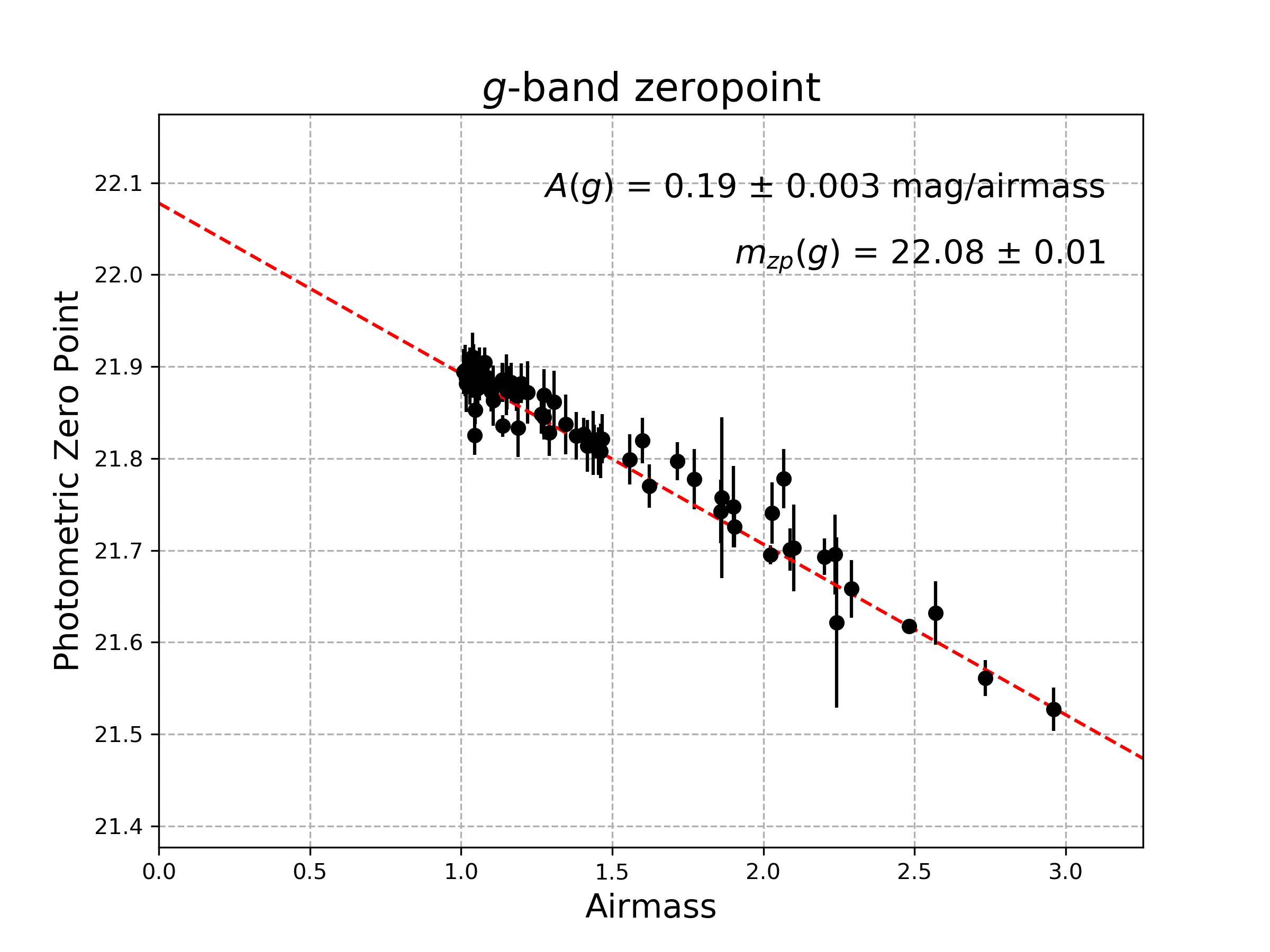}
    \caption{Photometric zero point as a function of airmass for all sky observations on UT 9 June 2020 in the $g$ band. }
    \label{fig:mzp}
\end{figure}

Figure~\ref{fig:mzp} shows the photometric zero point in the $g$ band as a function of airmass for our 9 June 2020 data. Eighty-nine of the 96 images were used for this plot. The remaining 7 images had tracking errors or insufficient Pan-STARRS sources to obtain a reliable photometric zero point. The data in the other bands is comparable, and the results are summarized in Table~\ref{tab:observatory}.

\subsection{First Order Color Correction}
\label{sec:color_correct}
Because our passbands do not match those of Pan-STARRS precisely, it is expected that our zero points may suffer small offsets based on the color of reference stars used. Using the same set of observations described in \S~\ref{sec:mzp}, we explored the first order color corrections of our magnitude scale using the Pan-STARRS catalog as a reference. 

For each image in each band, we matched the targets analyzed by DoPHOT against the catalog positions of Pan-STARRS sources in the field. For each pair of matched stars, a magnitude was derived using the counts derived by DoPHOT and Equation~\ref{eq:mag}. Then, the difference between the derived magnitude and the Pan-STARRS magnitude as well as the difference in the Pan-STARRS colors of the two stars were calculated. The magnitude differences and the color differences for all pairs in all images were compiled for further analysis.

\begin{deluxetable*}{@{\extracolsep{0.1in}}rl}
\tablenum{3}
\tablecaption{Pan-STARRS $griz$ Color Corrections\label{tab:color}}
\tablehead{Filter Band & Correction Equation}
\startdata
$g$\dotfill                         & $g_{corr} = (-0.035 \pm 0.005)\times\Delta(g-r) + g $ \\
$r$\dotfill                         & $r_{corr} = (0.034 \pm 0.002)\times\Delta(g-r) + r $ \\
$i$\dotfill                         & $i_{corr} = (-0.135 \pm 0.006)\times\Delta(i-z) + i $ \\
$z$\dotfill                         & $z_{corr} = (0.081 \pm 0.005)\times\Delta(i-z) + z $ \\
\enddata
\end{deluxetable*}

\begin{figure}
    \centering
    \includegraphics[width=1.0\columnwidth]{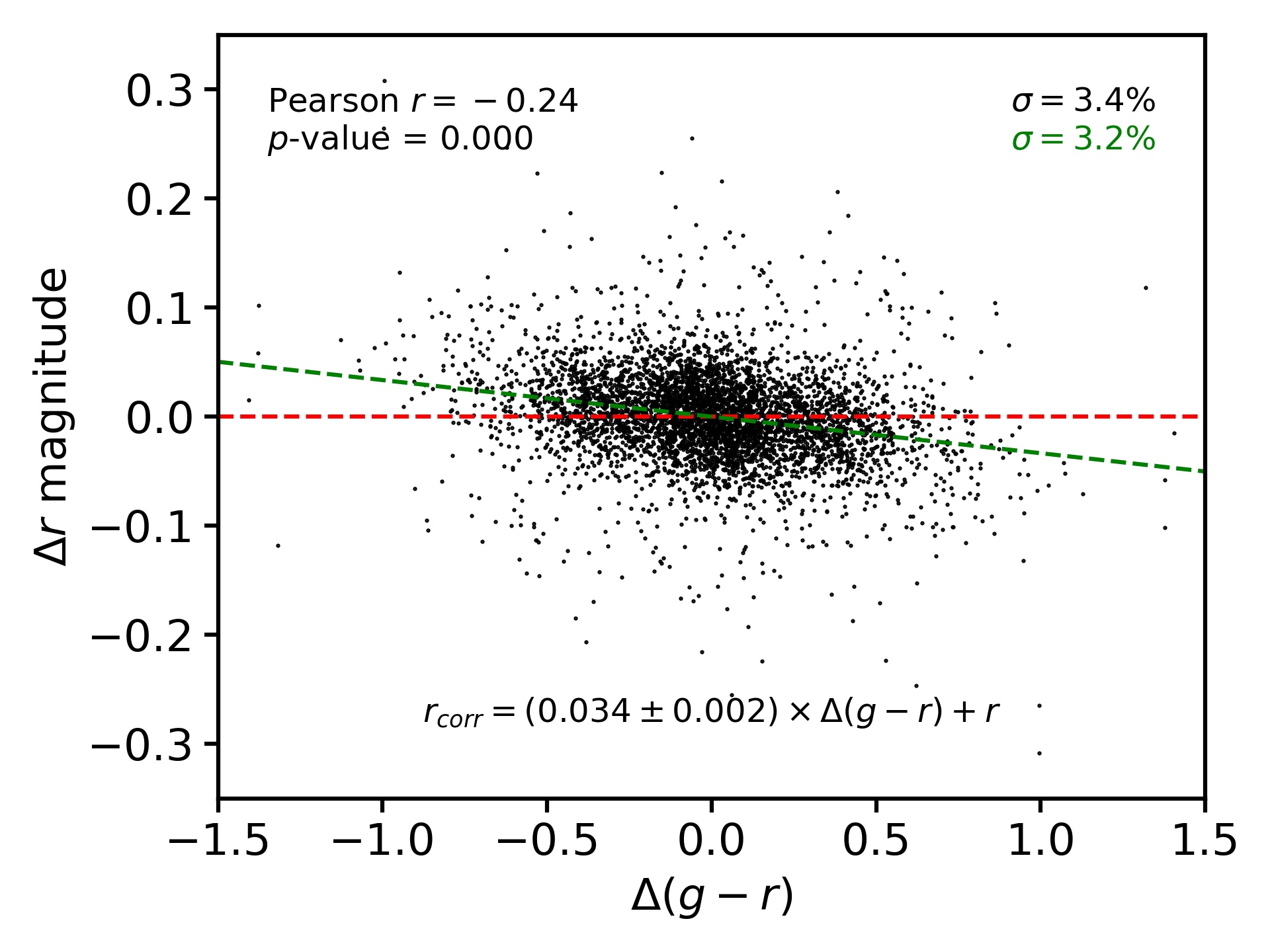}
    \caption{$r$-band color correction for our filter set as compared to the Pan-STARRS catalog. The data are derived from all-sky observations on a single, photometric night. }
    \label{fig:r_corr}
\end{figure}

\vspace{-24pt}
Figure~\ref{fig:r_corr} shows the difference in the derived $r$-band magnitudes and the Pan-STARRS magnitudes of all point source pairs in our all-sky data set as a function of the color difference $\Delta(g-r)$. This plot shows a clear trend in the data which suggests that for a bluer reference star the true magnitude of the target star will be brighter than the magnitude derived by using Equation~\ref{eq:mag} alone. This makes sense, as our $r^\prime$ filter cuts off redward of the Pan-STARRS $r$-band filter. This means that a blue reference star will produce more flux, and hence a lower magnitude, in our filer set. The opposite is also true: a redder reference star will produce a magnitude that is too high in our filter set and will need to be corrected. The color corrections for all our corresponding filters---$g^\prime$, $r^\prime$, $i^\prime$, and $z^\prime$---are outlined in Table~\ref{tab:color}.

\vspace{-12pt}
\subsection{Pointing and Tracking}
\label{sub:point_track}
\subsubsection{Pointing}
It is standard for our automation software to do a pointing correction before science observations commence using the latest version of the PinPoint Astrometric Engine\footnote{\url{http://pinpoint.dc3.com/}}. However, it is useful to know the accuracy of our dead reckoning pointing which can be assessed through the mount model used by the the telescope control software to convert between the celestial and horizon coordinate systems.

On the evening of October 31, 2018, we ran the mount model script built into PWI2 using $20^\circ$ degree steps in azimuth between $0^\circ$ and $360^\circ$ and $15^\circ$ steps between $12^\circ$ and $87^\circ$ for a total of 90 pointings over the entire sky. The model corrects the azimuth by fitting a function with 7 terms:
\begin{equation}
    \Delta \phi = \phi_0 + \sum_{N=1}^3 \left[ a_N\sin(N\phi) + b_N\cos(N\phi)\right]
    \label{eq:azcorr}
\end{equation}
where $a_N$ and $b_N$ are constant coefficents, and corrects the altitude by fitting a function with 9 terms:
\begin{eqnarray}
    \nonumber
    \Delta \theta = \theta_0 + c\sin(\theta)+d\cos(\theta) + \\
    \sum_{N=1}^3 \left[ f_N\sin(N\phi) + g_N\cos(N\phi)\right]
    \label{eq:altcorr}
\end{eqnarray}
where $c$, $d$, $f_N$ and $g_N$ are also constant coefficients.

The residuals of the pointing centers from the model are distributed isotropically with an RMS of $3.8^{\prime\prime}$ and a maximum deviation of $8.4^{\prime\prime}$. From these results we can expect our pointing to be better than $\sim10^{\prime\prime}$ with a typical error less than $\sim5^{\prime\prime}$. These numbers are consistent with our experience using the telescope.

\subsubsection{Tracking}
\label{subsub:tracking}
While the Thacher Observatory has an offset guide camera and our automation software can perform guided observations, it is our standard practice to simply track during short exposures because of the complexity and difficulty of getting offset guiding to work in an automated fashion. Our experience is that exposures shorter than 5 minutes show minimal signs of smearing or tracking errors. To validate this procedure, we performed a series of test observations on UT 25 and 26 August 2021. Six positions were chosen across the sky to represent a variety of observations---targets that were rising in the northeast, rising in the southeast, transiting the meridian near zenith in the south, transiting the meridian in the north, circumpolar, and setting in the west. Exposure times were all set to 10 seconds which is long enough to detect a large number of stars for plate solving and analysis of image quality, but short enough to track changes over the course of the hour long observations.

\begin{figure}
    \centering
    \includegraphics[width=\columnwidth]{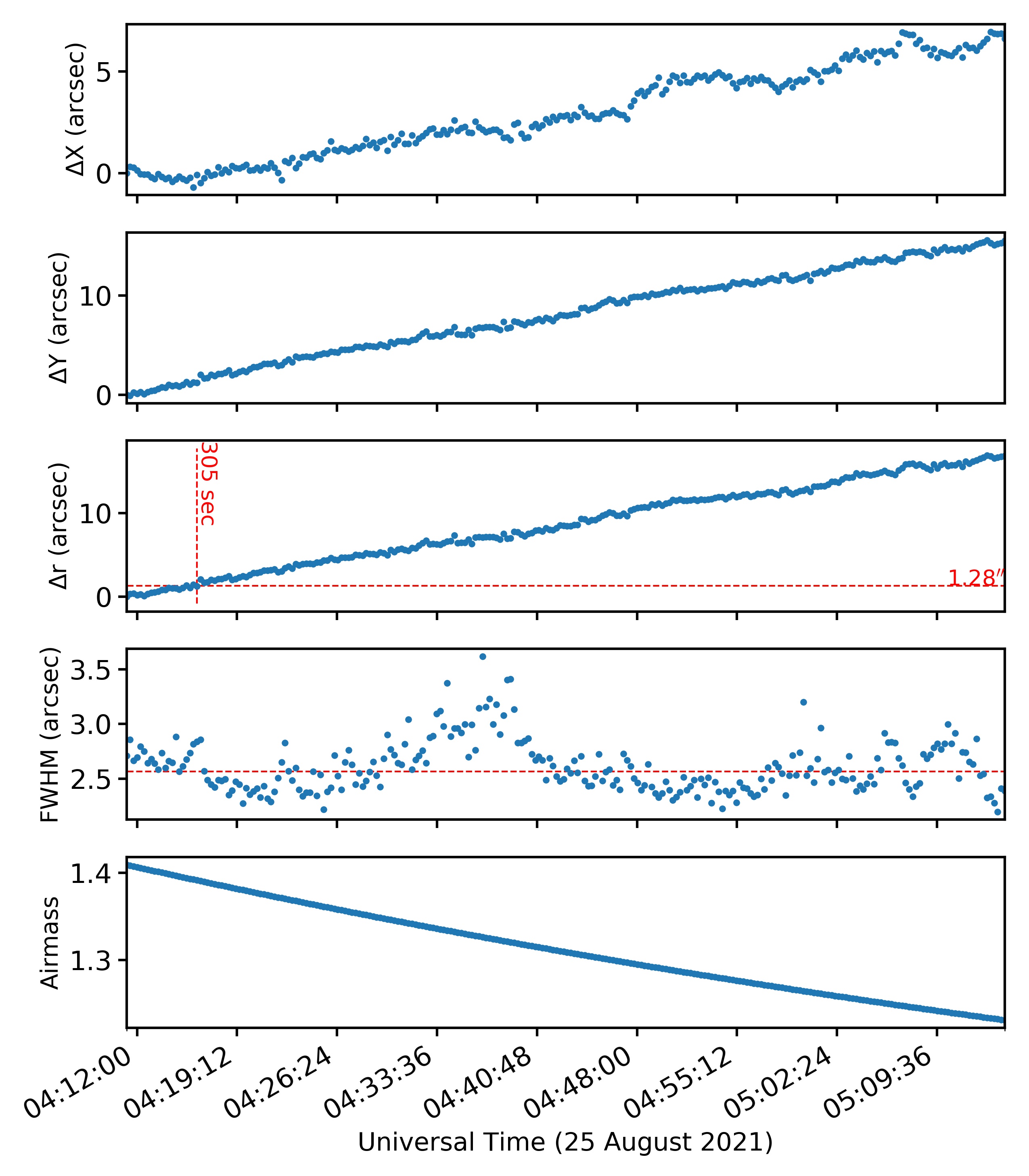}
    \caption{Results of one set of tracking observation made on UT 25 August 2021 consisting of 10 second exposures taken over the course of one hour starting at an altitude of $45^\circ$ and an azimuth of $45^\circ$. The median half width at half maximum (HWHM) of the point sources in the image is shown as the horizontal red dashed line in the middle ($\Delta r$) panel---this is half of the median FWHM shown as the horizontal red dashed line in the FWHM panel. The time that it took for the center of field to drift a total angle equal to the median HWHM is shown as the vertical red dashed line. This represents the longest exposure time that can be used without significant smearing for un-guided observations.}
    \label{fig:tracking}
\end{figure}

Figure~\ref{fig:tracking} shows the results of one of our observations made on UT 25 August 2021 starting at an altitude of $45^\circ$ and an azimuth of $45^\circ$. The images were analyzed using SExtractor \citep{Bertin1996}, and only point sources above a signal to noise ratio of 10 were considered to derive a median FWHM of each image. We used \texttt{astrometry.net} \citep{Lang2010} to plate solve each image, and the celestial coordinates of the center pixel of the first image ($\alpha_0,\delta_0$) was determined using utilities in the \texttt{astropy.wcs} \citep{Astropy1-2013,Astropy2-2018} package. The pixel value of ($\alpha_0,\delta_0$) was then tracked for each image and the offset from the center pixel calculated in arcseconds using a tangent projection.

The other five tracking observations showed quite a bit of variance in these results, with the mean time until a drift of one median HWHM ranging from 153 to 520 seconds. The tracking experiments with the worst tracking were the circumpolar target---which started $6^\circ$ above the north celestical pole---and the setting target in the west---which started at an azimuth of $260^\circ$ and an altitude of $45^\circ$. The best tracking happened for the target rising in the southeast starting from an azimuth and altitude of $115^\circ$ and $30^\circ$ respectively. This can be partially explained by the poorer image quality at low elevation. If we do not consider the median HWHM of the images and instead use the typical HWHM of $1.5^{\prime\prime}$, we find that the tracking of all observations remain with this absolute angle for more than 5 minutes. 

We found the tracking drift to be approximately linear, with a mean drift rate of 0.322 arcseconds per minute. Our characteristic seeing is $3^{\prime\prime}$, so that over $\sim 5$ minutes the telescope will drift about one HWHM.

\subsection{Timing}
\label{sec:timing}
Since many of our scientific goals involve precise and accurate timing, we have acquired a Stratum 1 NTP time server from Masterclock, Inc. This unit is a GPS referenced NTP time server that comes with a preamplified antenna. From November, 2018 until early January 2022 we had been using the Dimension 4 software\footnote{\url{http://www.thinkman.com/dimension4/}} to reference and apply corrections to our Windows 10 software every minute. 

\begin{figure}
    \centering
    \includegraphics[width=1.0\columnwidth]{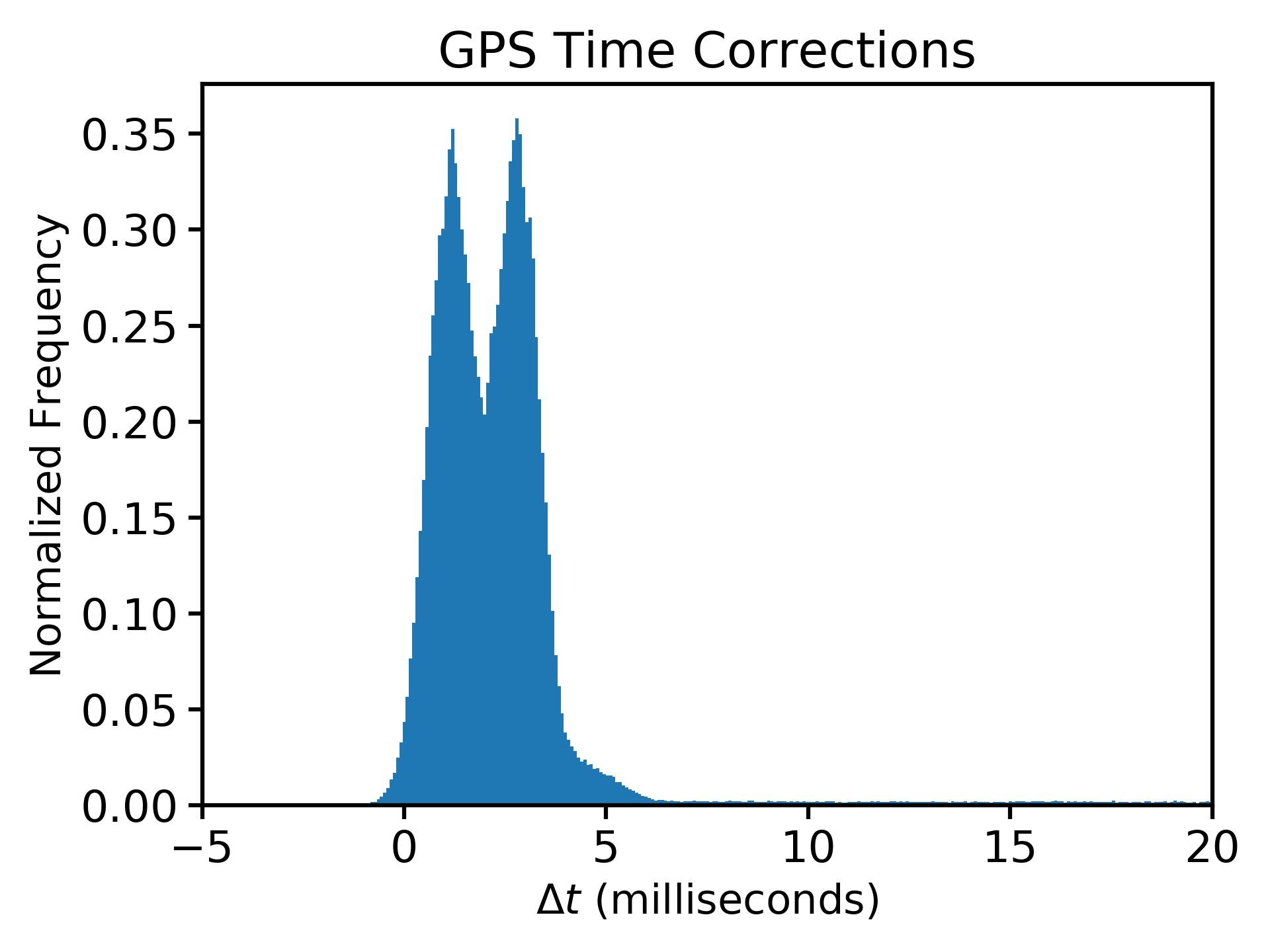}
    \caption{Histogram of time corrections applied to our control computer each minute between November 1, 2018 and June 9, 2021. The full range of corrections is larger than shown---see text for details. However, the frequency of corrections beyond the shown range are so small as to render them invisible on this scale.}
    \label{fig:dt}
\end{figure}

From 2.6 years of nearly continuous data, we find the median correction applied to our Windows 10 operating system to be $+4.1$\,ms. Unexpectedly, the Windows software was found to be lagging behind the GPS server 99.3\% of the time, and the distribution of corrections is highly skewed and double peaked with the 1-$\sigma$ interval for the corrections being from 1\,ms to 0.34\,s. The full range of corrections span from -8.5 seconds up to 1.5 seconds. However, corrections with absolute magnitude larger than 1 second are vanishingly rare (only 25 out of 1.4 million corrections). The main clump of time corrections is shown in Figure~\ref{fig:dt}. The very low level of corrections larger than $\sim 6$\,ms that stretch out to $\sim 1$ second are not easily visible in this plot due to the linear scale. 

It is also important to consider system latency when assessing the accuracy of our timing. The latency of USB 2.0 and CAT5 cable is approximately 5\,ns per meter. The cable length between the control room and our camera is less than 10 meters adding less than 0.05\,ms latency to our system. The technical specifications on our Black Box USB extenders quote less than 1\,ms of latency. So, conservatively, we can be confident that the timing of our observations is better than 0.5 seconds. However, the accuracy is more typically in the millisecond range. For any time sensitive observations, the timing accuracy can be directly accessed through the history of Dimension 4 time corrections during the observations.

After extensive testing with a local time server, we found that Dimension 4 was not consistent and showed 40 millisecond jumps in consecutive syncs even when the ping latency to the server was under 1 millisecond. We are unsure about the causes of this behavior, but it may explain the bimodal distribution seen in Figure~\ref{fig:dt}. At the time of the writing of this paper, we have switched to using NetTime\footnote{\url{https://www.timesynctool.com/}} to update our system every 10 minutes and we have seen more consistent results. Testing of this tool suggests that our observatory timing is now accurate to 5 milliseconds. 

\section{Data Reduction and Analysis}
\label{sec:science}
We run our observations in an automated fashion using the ACP Expert Scheduler software. Bias and dark calibration frames are performed every morning following observations and flat field observations are run manually on approximately a monthly basis. A 750\,mm Aurora Flat Field Panel has been recently purchased which can be automated. Following its full installation, flat fields are expected to be performed every evening.

Photometry is performed by one of two different general methods depending on the specifics of the field in question and the scientific requirements. While our pipeline is evolving and is subject to change, we outline the basic features of each method as our data has been used in published scientific works including the monitoring of Boyajian's Star \citep{Boyajian2018,Wilcox2019} and Boyajian Star analogs \citep{Arculli2020,Browning2021}, eclipsing binaries \citep{Healy2019}, transiting exoplanets \citep{Jin-Ngo2021}, and transient studies \citep{Tinyanont2021,Sand2021,Kilpatrick2021,Gagliano2021,Barna2021,Jacobson-Galan2020,Swift2019,Swift2018}

\begin{figure}[!t]
    \centering
    \includegraphics[width=1.0\columnwidth]{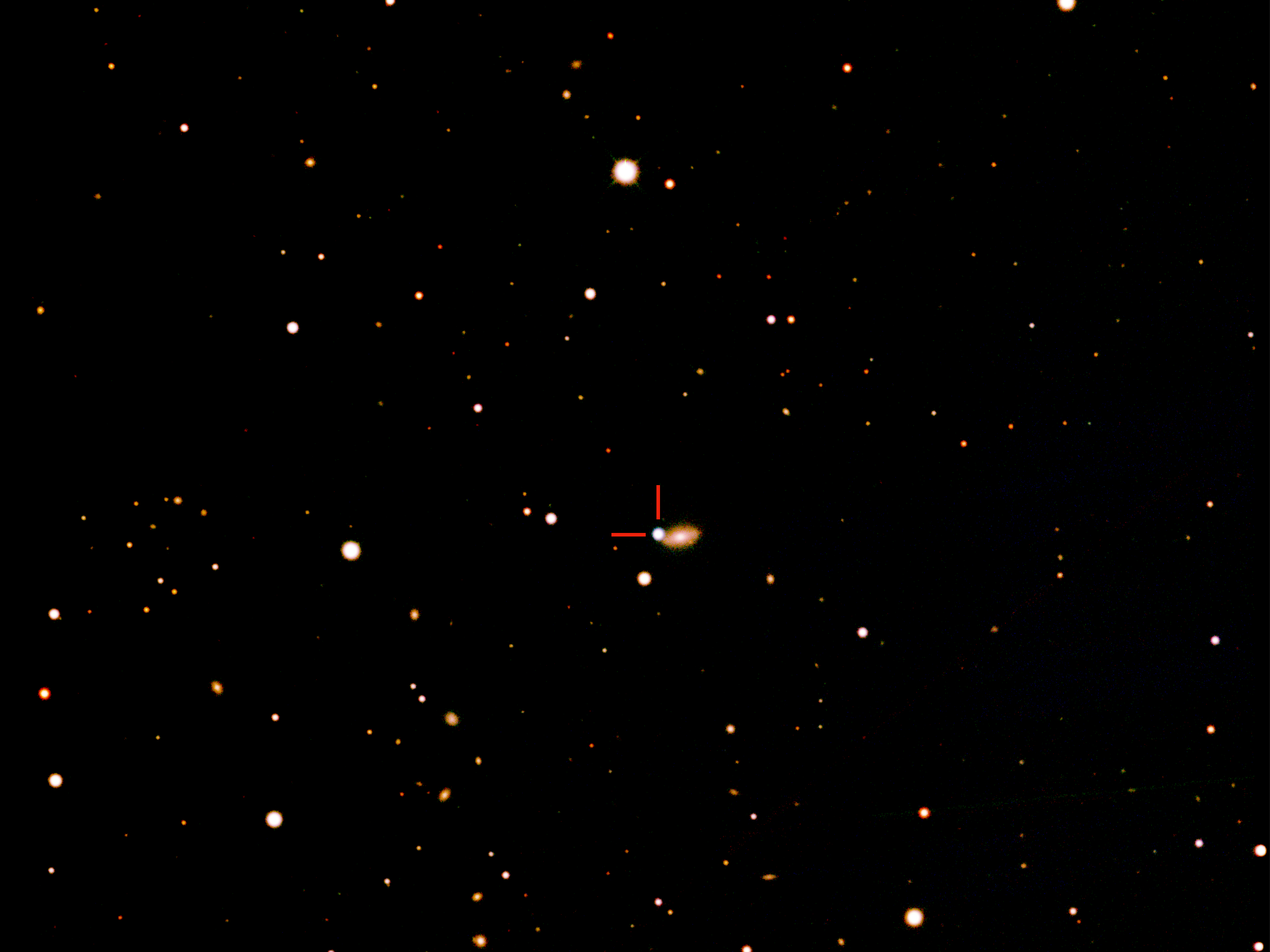}
    \caption{Full frame composite image of the SN\,2020hvf field constructed with stacked images in the $g$, $r$, and $i$ photometric bands. The marker indicates the supernova. }
    \label{fig:20hvf_image}
\end{figure}

\subsection{Preliminary Data Handling and Calibrations}
\label{sec:cal}
Each observation is stored as a file according to the Flexible Image Transport System (FITS) format in a staging area within our institution's Drop Box. A running script on our local network checks this staging area each morning and performs several preliminary functions on our data before moving them to our data archive---also on Drop Box.

First, while each observation is typically plate solved through ACP with PinPoint, which utilizes ATLAS-REFCAT 2 \citep{Tonry2018}, a secondary attempt to plate solve images that either failed or otherwise were not solved using a local version of \texttt{astrometry.net} \citep{Lang2010}. Log files, plate solved science files, calibration frames as well as master bias and dark frames, and all-sky images converted into an MPEG-4 video file are moved to the data archive in a directory named according to the UT date of the observations.

The data in the archive are not changed or manipulated. However, before further analysis, each observation is calibrated in the standard way: 
\begin{equation}
C = \frac{R - B - D\times t}{F}
\label{eq:cal}
\end{equation}
where $C$ is the calibrated frame, $R$ is the raw image, $B$ is the master bias frame, $D$ is the bias subtracted master dark frame normalized to be in units of ADU/s, and $F$ is the master flat frame in the appropriate photometric band. Master frames are created using a median filter created from 30 or more individual frames. If calibration frames are not available from the night of observation for whatever reason (\textit{e.g.}, hardware or software failure), master calibration frames from the archive are used.

\begin{figure*}[!ht]
    \centering
    \includegraphics[width=6.5in]{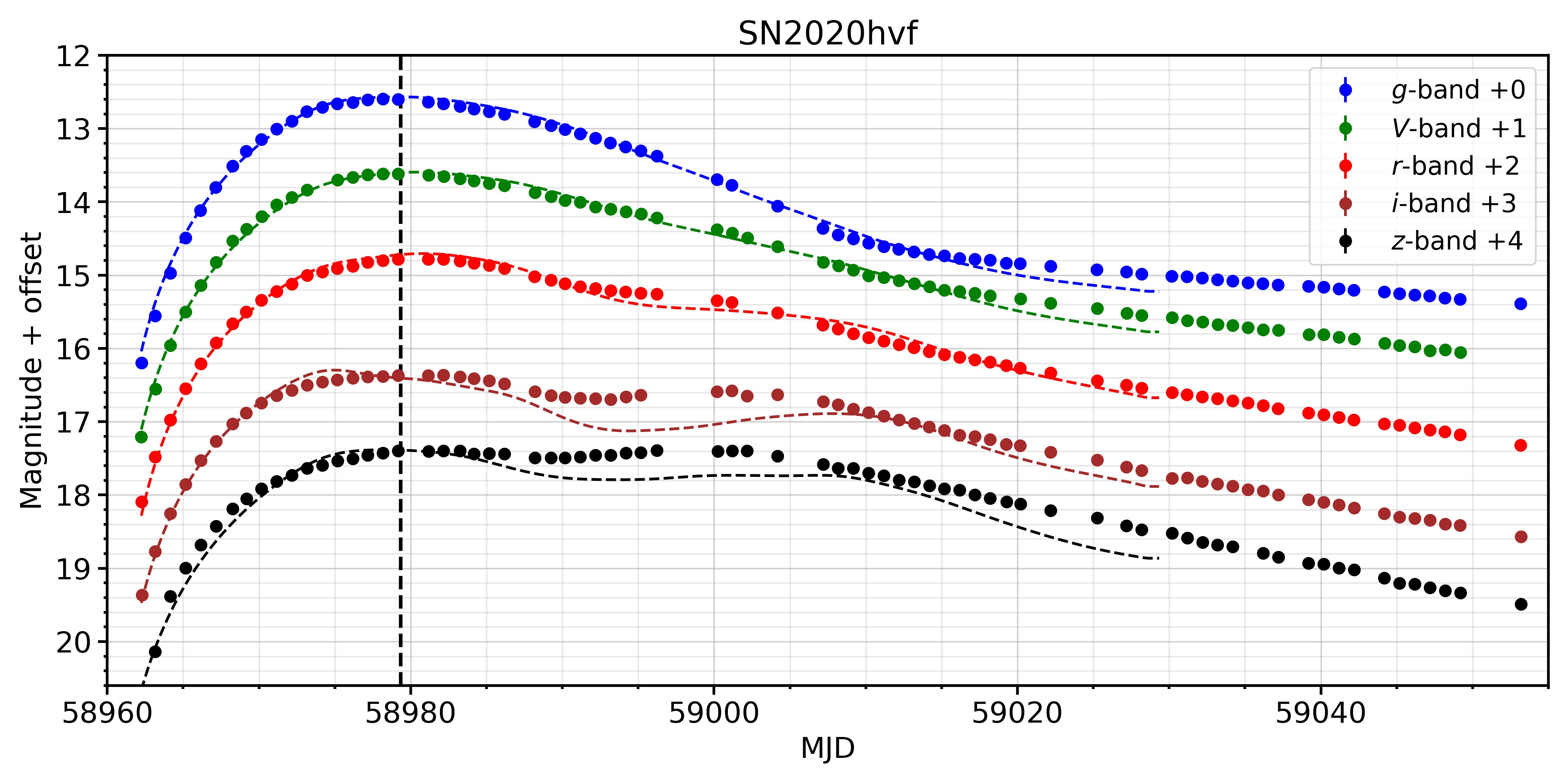}
    \caption{Multiband lightcurve of SN\,2020hvf. Data reduction and photometry were performed according to \S\ref{sec:cal} and \S\ref{sec:psf}. Counts were obtained using PSF fitting and magnitudes were determined via comparison with the Pan-STARRS catalog. Errors are shown, but fall mostly within the data points.}
    \label{fig:20hvf_lc}
\end{figure*}

\subsection{PSF Fitting Photometry}
\label{sec:psf}
For science targets that are in crowded fields or sit atop a background (\textit{e.g.}, supernovae), we employ PSF fitting photometry using DoPHOT \citep{Schechter1993}. We use SN\,2020hvf \citep{Smith2020} as an example for our PSF photometry, and we present the full, multi-band light curve in Figure~\ref{fig:20hvf_lc}.

SN~2020hvf is a nearby type Ia supernova \citep{Burke2020} located in galaxy NGC\,3643 \citep{Smith2020} which has a redshift of 0.00581 \citep{Bolton2012} and a distance modulus of $32.45\pm0.15$~mag \citep{Jiang2021}. It is a carbon-rich, high-luminosity type Ia supernova that may be the result of a ``super-Chandrasekhar'' thermonuclear explosion (where the total mass is larger than the Chandrasekhar mass). Early, short-lived excess emission is explained as the interaction between the supernova ejecta and $\sim 0.01\,M_\odot$ of circumstellar material extending out to $\sim 1$\,AU \citep{Jiang2021}. 

This supernova has many attributes that make it a good candidate for followup observations at the Thacher Observatory: it is bright, it is located in a region with a relatively simple background, and has a declination $85^\circ \gtrsim \delta \gtrsim -15^\circ$. It was also discovered early \citep{Smith2020} and was observable for approximately 3 months before setting. Observations of SN~2020hvf were performed in all five of our wide photometric bands ($griz$, and $V$). Integration times in all bands were started out at 120\,s, then decreased to 90\,s near the peak of the light curve to prevent saturation, then increased again to 120\,s once the light curve was on the decline. The varying exposure times allowed us to achieve typical signal to noise ratios of 250, 350, 400, 300, and 130 in $g$, $V$, $r$, $i$, and $z$, respectively. Our first observation was on UT April 23, after the ``excess'' emission had faded, and we continued nearly nightly monitoring for 90 subsequent evenings. The last observation was on UT July 23. 

Each observation of SN\,2020hvf consisted of 3 separate exposures in each photometric band. On some nights, the target was observed twice for a total of 6 exposures per band per night. After standard calibrations (\S\ref{sec:cal}), Source Extractor \citep{Bertin1996} was run on each image to obtain preliminary image information and statistics, including the FWHM of point sources and the sky background level. These statistics were then used to create a DoPHOT parameter file customized to the image which helps with efficiency and convergence.

Counts for the target of interest are compared to the counts obtained for Pan-STARRS sources \citep{Flewelling2020,Magnier2020} in the field that have a maximum error on the reported photometry of 2\%, a maximum standard deviation of 2\%, and lie between magnitudes of 12 and the $10\,\sigma$ detection limit of the image. This resulted in 14 stars 
that were used to derive zero points in all photometric bands using magnitudes from the Pan-STARRS catalog. The counts from each reference star were used to convert the target counts into magnitudes on the Pan-STARRS photometric scale using Equation~\ref{eq:mag}. Errors on the magnitudes are calculated by propagating the error on the flux as reported by DoPHOT and then adding the reported standard deviation and error on the magnitude of the reference star in quadrature. A weighted mean of the magnitudes derived from each reference star along with their associated errors is used to determine a preliminary magnitude and magnitude error in each image.

A second iteration is made in the data reduction process to perform color corrections. During this pass the preliminary magnitudes and errors are used to estimate source colors. The same procedure as described above is then applied to the data. However, the magnitudes of the target derived from each reference star is corrected according to the equations in Table~\ref{tab:color}. The weighted mean of the color-corrected magnitudes derived from each reference star along with their associated errors is used to determine a final, color-corrected magnitude and magnitude error. A single photometric measurement is then calculated for each night by a weighted mean of all individual measurements derived. We also apply a magnitude error floor of 0.02 mags that corresponds approximately to the mean RMS between our derived reference star magnitudes compared to the the Pan-STARRS catalog magnitudes. Figure~\ref{fig:20hvf_lc} shows the results of this reduction applied to our SN~2020hvf data set.

We fit our SN~2020hvf photometry with the SALT3 model \citep{Kenworthy2021} using the python package \texttt{sncosmo} \citep{Barbary2021}\footnote{\url{https://sncosmo.readthedocs.io/}}.  This model is used to derive distances to Type Ia supernovae for cosmological studies, but can also provide basic information about a supernova.  When fitting, we fixed the redshift to that of the host galaxy, $z=0.00581$; the remaining four parameters, $t_0$ (the time of peak $B$-band magnitude), $x_0$ (the flux scale parameter), $x_1$ (the light-curve stretch parameter),  and $c$ (the phase independent color term), are allowed to float.

Fitting all bands, the best-fitting SALT3 model could not reproduce the flux in the redder bands after peak light, with SN~2020hvf generally being brighter than the model in $i$, and $z$ at those times. In particular, it looks like the second-maxima for SN~2020hvf are brighter and peak earlier than the best-fitting SALT3 model.  We then refit the data, restricting to the $gVr$ bands. The best-fit parameters, as determined by the built in \texttt{fit\_lc} method in \texttt{sncosmo}, are shown in Table~\ref{tab:snfit}. These best-fit parameters were then used to create model light curves for SN~2020hvf in all photometric bands (dashed lines in Figure~\ref{fig:20hvf_lc}). The model fits the bluer bands reasonably well, but the differences in the redder bands remain. The light-curve parameters are within the normal bounds of cosmological samples of Type Ia supernovae. We therefore suspect that the difference between SN~2020hvf and the SALT3 model is related to its peculiarities as described by \citet{Jiang2021}.

SALT3 indicates that SN~2020hvf had a peak $B$-band luminosity on 58979.339 MJD, consistent with the peak $B$-band brightness derived by \citet{Jiang2021} from a polynomial fit to their $B$-band light curve\footnote{Note that we did not observe SN~2020hvf in the $B$ band and thus cannot compare to a direct measurement of our own photometry.}. Meanwhile the stretch parameter, $x_1$, is positive (0.809) reflecting the long rise time and small $\Delta m_{15}(B)$ measured by \citet{Jiang2021}.  If SN~2020hvf follows the \citet{Phillips1993} relation, this would indicate a high-luminosity supernova. Using the SALT3 model $B$-band light curve and the distance modulus of SN~2020hvf from \citet{Jiang2021}, we derive a peak $B$-band absolute magnitude of $-19.88$, also consistent with the value derived by \citet{Jiang2021}. The best-fitting color term, $c$, suggests that SN~2020hvf is slightly bluer than the mean of the SALT3 training sample, and likely had minimal host-galaxy reddening.

\begin{deluxetable}{@{\extracolsep{0.25in}}rl}
\tablenum{4}
\tablecaption{SALT3 Model Fit to SN~2020hvf\label{tab:snfit}}
\tablehead{Parameter & Best Fit Value}
\startdata
$z$ (fixed)\dotfill  & $0.00581$ \\
$t_0$ (MJD) \dotfill & $58979.339 \pm 0.017$ \\
$x_0$ \dotfill       & $0.16429 \pm 0.00082$ \\
$x_1$ \dotfill       & $0.809 \pm 0.024$\\
$c$ \dotfill         & $-0.0494 \pm 0.0044$ \\
\enddata
\end{deluxetable}

\subsection{Aperture Photometry}
\label{sec:aperture}
To demonstrate the viability of our facility to obtain precision light curves via aperture photometry, we observed a transit of the hot-Jupiter Qatar-1b \citep{Alsubai2011}. Qatar-1 is a $V = 12.84$ metal-rich K dwarf star located at high declination near the Galactic plane. We observed Qatar-1b on the night of UT 1 October 2021 in the $r^\prime$-band using 90 second integration times. The observations started at UT 04:07:40, ended at UT 11:35:14 and consisted of 250 individual images, all of which were used to create a light curve.

\begin{deluxetable*}{@{\extracolsep{0.1in}}rl}
\tablenum{5}
\tablecaption{Qatar 1b Transit Fit\label{tab:transit}}
\tablehead{\multicolumn{2}{c}{Fixed Parameters}}
\startdata
$P$: Period (days) \dotfill & 1.4200243 \\
$e$: Eccentricity \dotfill & 0 \\
\hline 
\multicolumn{2}{c}{Fit Parameters} \\
\hline 
$t_0$: Mid-transit time (BJD)\dotfill & $2459488.79915^{+0.00040}_{-0.00011}$\\ 
$R_p/R_\star$: Scaled planet radius\dotfill & $0.1435^{+0.0001}_{-0.0056}$ \\
$a/R_\star$: Scaled semi-major axis\dotfill & $6.40^{+0.77}_{-0.21}$ \\
$i$: Orbital inclination (degrees)\dotfill & $84.9^{+1.8}_{-0.4}$ \\
$q_1$\tablenotemark{a}: First limb darkening parameter\dotfill & $0.47^{+0.37}_{-0.17}$ \\
$q_2$\tablenotemark{a}: Second limb darkening parameter\dotfill& $0.35^{+0.19}_{-0.24}$  \\
\hline 
\multicolumn{2}{c}{Derived Parameters} \\
\hline 
$\delta$: Transit depth \dotfill & 0.0227 \\
$T_{14}$: Full transit duration (hours) \dotfill & 1.68 \\
$\rho_\star$: Stellar density (cgs) \dotfill & $3.06^{+0.79}_{-0.68}$ \\
\enddata
\tablenotetext{a}{\citet{Kipping2013a}}
\end{deluxetable*}

\begin{figure}[!hb]
    \includegraphics[width=1.05\columnwidth]{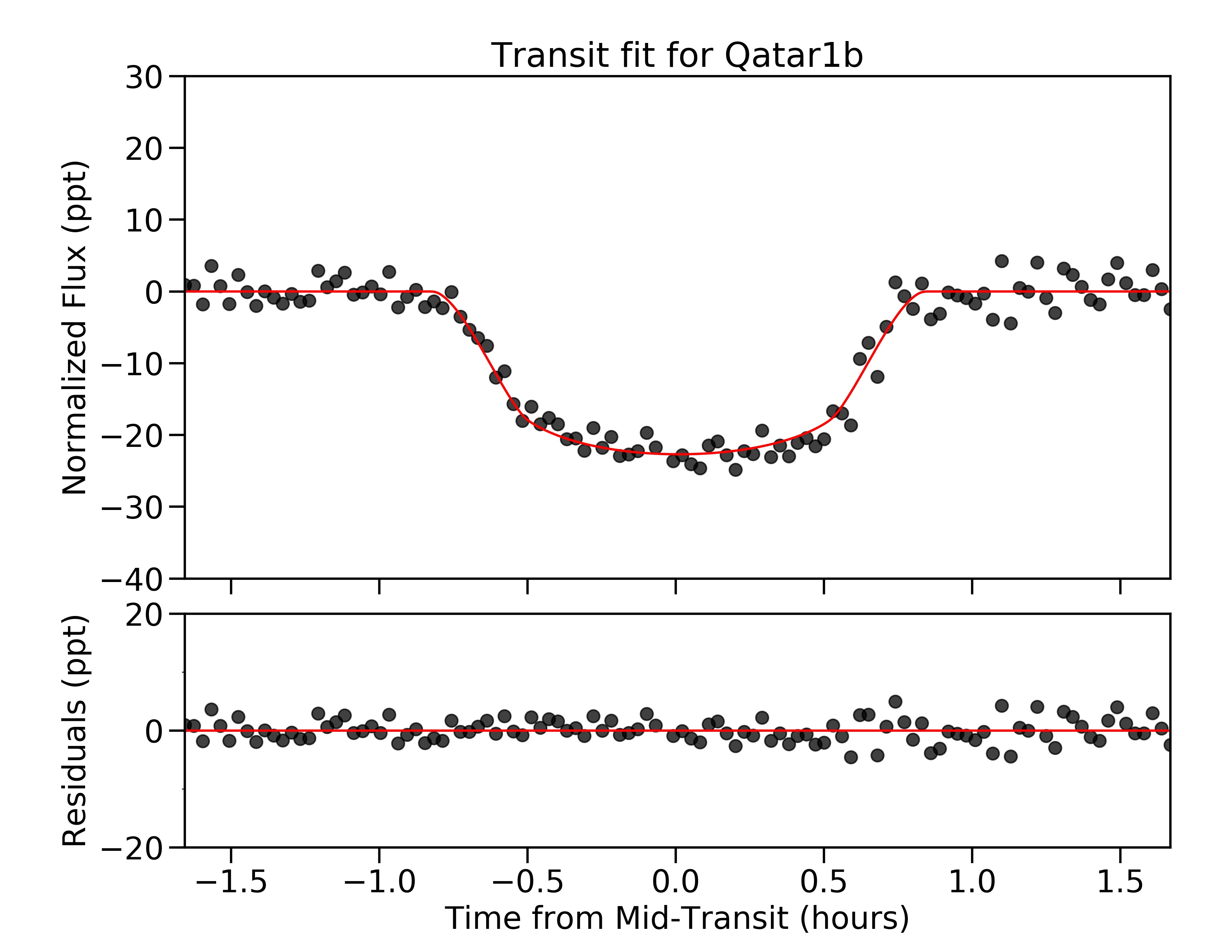}
    \caption{Transit light curve of Qatar-1b taken on the evening of UT 1 October 2021. The best fit model determined by our analysis and the residuals are shown. The RMS of the residuals is 1.6\,mmag.}
    \label{fig:transit}
\end{figure}

\vspace{-24pt}
The 250 images were calibrated and stacked, reference stars were chosen by hand, and the radii of the sky annulus for each star as well as any aperture restrictions (\textit{e.g.}, due to neighboring stars) were determined. Each image was analyzed to compute the aperture that would produce the highest signal to noise for Qatar 1. This aperture was used to measure counts in the target and reference stars using the \texttt{photutils} package \citep{Bradley2020}. Six reference stars were used to calculate the final differential photometry. A single data point was excluded from the resultant light curve as it was a clear outlier. Then, a slight trend was fit as a function of airmass for the out-of-transit data using a line, and this fit was then divided into the data to produce the final light curve shown in Figure~\ref{fig:transit}.

\begin{figure*}
    \centering
    \includegraphics[width=6in]{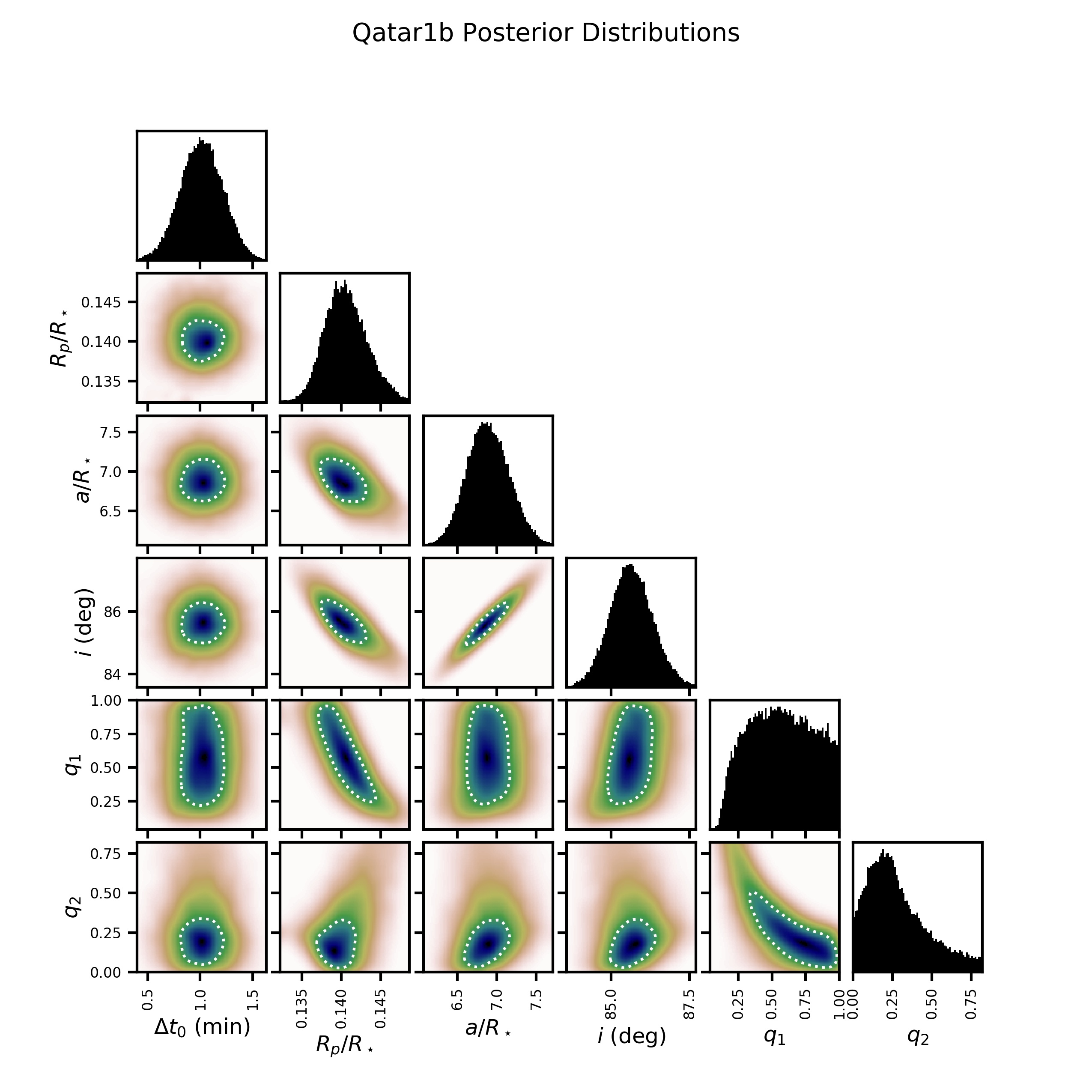}
    \caption{Posterior probability distribution for a transit model applied to our Qatar 1b observations. The 68\% confidence contour for each 2-D slice of the posterior is shown as the dotted white line.}
    \label{fig:posteriors}
\end{figure*}

A best fit transit model was generated using \texttt{batman} \citep{Kreidberg2015} to calculate model light curves for a given sets of input parameters. Reference values for our transit model parameters were derived from the planet and star properties reported by \citet{Alsubai2011} and \citet{Collins2017}, and then the posterior probability distribution for the free parameters $R_p/R_\star$, $a/R_\star$, $i$, $t_0$, and $q_1$ and $q_2$ quadratic limb darkening parameters \citep{Kipping2013a}, were sampled using the affine invariant Markov chain Monte Carlo (MCMC) ensemble sampler \texttt{emcee} \citep{Foreman-Mackey2013}. In Figures~\ref{fig:transit} and \ref{fig:posteriors}, the mid transit time, $t_0$ is expressed as Barycentric Julian Day (TDB) relative to the ephemeris given on the Swarthmore Transit Finder site\footnote{\url{https://astro.swarthmore.edu/transits/transits.cgi}}, BJD 2459488.7987. 

To sample the posterior, we first used 250 walkers and 10000 steps and analyzed the chains for convergence using the auto-correlation length. The sampler was then re-started to produce fully converged samples with an acceptance fraction of 0.43. The maximum value of the posterior probability was used to determine the best fit model parameters and the best fit model light curve shown in Figure~\ref{fig:transit}. The residuals are well behaved showing a slight increase in scatter with time due to increasing airmass. The RMS of the residuals is 1.6\,mmag. A triangle plot of 1-D and 2-D marginalized posterior probability distributions derived from our MCMC analysis is shown in Figure~\ref{fig:posteriors}.

Our final transit parameter values and $1\,\sigma$ intervals are reported in Table~\ref{tab:transit}. We find the transit timing to be consistent with the value on the Swarthmore Tranist Finder within about 1-$\sigma$ (stated error on the mid-transit time is $\pm 1$ minute). We also find the timing to be reasonably consistent with all previous studies, individually \citep{Alsubai2011,Covino2013,vonEssen2013,Mislis2015,Maciejewski2015,Cruz2016,vonEssen2017,Bonomo2017,Collins2017}. Two notable exceptions are the predictions from \citet{Alsubai2011} and \citet{Mislis2015} who published ephemerides that were 2.0-$\sigma$ early and 2.9-$\sigma$ late, respectively, in comparison to our measured transit time. Our derived values for all other quantities in our transit fit are largely in accord with published values, and while the limb darkening parameters were not particularly well constrained by our data, our best fit values for the limb darkening parameters are close to what is expected from stellar atmosphere models: $u_1 = 0.628$ $u_2 = 0.104$ from \citet{Claret2011} or equivalently $q_1 = 0.536$ and $q_2 = 0.429$ \citep{Kipping2013a}. 

\section{Concluding Remarks}
\label{sec:conclusions}
The Thacher Observatory has a legacy dating back to the late 1950's and was fully renovated into a state-of-the-art facility in 2016. The observatory now houses a 0.7-m PlaneWave telescope, a new 16.5-foot Ash Dome, new control and weather computers, and a suite of software that allows it to be run in a fully automated fashion. Located at the Thacher School in Ojai, CA, which abuts the Los Padres National Forest, the site affords dark skies, $>270$ clear nights per year on average, and $\sim 3^{\prime\prime}$ seeing.

Detailed information about the site, descriptions of our hardware and software, and the full capabilities of the observatory have been outlined to both provide background and insight into the data from the Thacher Observatory that has already been included in the literature, as well as to promote further collaborations and continued contributions to the scientific community moving forward.

The stated observatory capabilities have also been demonstrated through two separate studies that highlight both our PSF fitting photometry that is well suited for supernovae and stellar photometry in crowded fields, and standard aperture photometry. A light curve of SN2020hvf is presented that includes nearly nightly photometry spanning 75 days in 5 photometric bands. We also present a transit observation of Qatar-1 b that reproduces the accepted parameters of this well-studied target. 

\begin{acknowledgments}
The fully renovated Thacher Observatory would not have been possible were it not for the trust, support, and shared vision by many mentors, colleagues, and donors. We are truly thankful for the opportunity provided by the Thacher School administration, development, and donors for helping make this observatory a reality.

We are indebted to John A. Johnson, Philip Muirhead, Lucianne Walkowicz, Ben Montet, Yutong Shan, and Eunkyu Han for their support, freely shared ideas, and the opportunities extended to our students which have helped us remain active in astronomical research. The collaboration and mentorship offered to Thacher students by Jason Eastman and the MINERVA team were gracious and foundation building for our Astronomy Program, and we have benefited greatly from Las Cumbres Observatory's dedication to education, outreach and willingness to invest in young minds
 
The UCSC team is supported in part by NASA grants NNG16PJ34C and NNG17PX03C; NSF grants AST-1518052, AST-1815935, and AST-1911206; the Gordon and Betty Moore Foundation; the Heising-Simons Foundation; and by a fellowship from
the David and Lucile Packard Foundation to R.J.F.

JJS would also like to thank Jack Welch for many years of support and guidance, Jill Tarter for her wisdom and vision, and the courage, foresight, and integrity  of John A. Johnson.

\end{acknowledgments}

\facility{Thacher Observatory}

\software{\texttt{astrometry.net} \citep{Lang2010}, \texttt{SExtractor} \citep{Bertin1996}, \texttt{DoPHOT} \citep{Schechter1993}, \texttt{astropy} \citep{Astropy1-2013,Astropy2-2018}, \texttt{sncosmo} \citep{Barbary2021}, \texttt{photutils} \citep{Bradley2020}, \texttt{emcee} \citep{Foreman-Mackey2013}, \texttt{batman} \citep{Kreidberg2015}}

\bibliography{_references}{}
\bibliographystyle{aasjournal}


\end{document}